\documentclass[aps,pra,twocolumn,groupedaddress,longbibliography]{revtex4-2}

\usepackage{amsmath}
\usepackage{amssymb}
\usepackage{graphicx}

\usepackage{mathtools}
\newcommand{\RNum}[1]{\uppercase\expandafter{\romannumeral #1\relax}}

\begin{document}

%\preprint{}

\title{Qubit entanglement generated by classical light driving an optical cavity}


\author{Seongjin Ahn}
%\thanks{}
\author{Andrey S. Moskalenko}
\email[]{moskalenko@kaist.ac.kr}
\affiliation{Department of Physics, KAIST, Daejeon 34141, Republic of Korea}

\author{Vladimir Y. Chernyak}
\email[]{chernyak@chem.wayne.edu}
\affiliation{Department of Chemistry, Wayne State University, 5101 Cass Ave, Detroit, Michigan 48202, USA}
\affiliation{Department of Mathematics, Wayne State University, 656 W. Kirby, Detroit, Michigan 48202, USA}

\author{Shaul Mukamel}
\email[]{smukamel@uci.edu}
\affiliation{Department of Chemistry, University of California, Irvine, California 92614, USA}


\date{\today}

\begin{abstract}
	We study the generation of entanglement between two qubits which communicate through a single cavity mode of quantum light but have no direct interaction. 
	We show that such entanglement can be generated simply by exchanging quanta with a third party, which is in our case the cavity mode. Exchanging only a single quantum creates maximal entanglement. A single quantum can be provided by an external quantum light source. However, we use a classical light source to pump quanta which are used for the exchange, and investigate the degree of two-qubit entanglement. 
	We first identify a characteristic timescale of the interaction between the cavity mode and each qubit. We investigate two regimes of the driving pulse length, one is short and the other is long compared to the characteristic timescale of the interaction.
	In the first regime, it is known that the pulse can pump the system by generating a displacement of the cavity mode. 
	We show that, by using a specific pulse shape, one can make the displacement to essentially vanish after the pulse finishes interaction with the cavity mode. In this case, a rotation of the qubits can be invoked. In addition, higher-order effects of the pulse including a non-local operation on the joint system of the cavity mode and the qubits are found, and we present a formalism to compute each term up to a given order. An explicit condition on the pulse shape for each term to be nonzero or suppressed is derived to enable an experimental design for verifying the entanglement generation using a classical light source. In the opposite regime where the driving is sufficiently long, we utilize a squeezed state which may be obtained adiabatically. We study how the squeezing and the accompanied rotation of qubits affect the generated two-qubit entanglement.
%	We show how the generated entanglement depends on the pulse strength, duration and the temporal shape. We demonstrate that the largest entanglement is obtained by a moderate pulse strength rather than observed in the strong driving limit. 
%	We explain the reason by comparison with the case where the cavity mode is prepared in a Fock state. 
%	We further show that this explanation does not hold when the driving becomes strong enough where there arises nonvanishing entanglement whereas the Fock state with a large number of photons cannot produce entanglement effectively.
%	Especially, we show that the 
%	Especially, classify the pulse shape 
	
%	[implication] classical light source -- available -- also, it can be used two objects in a separate .. direction need not be limited to directi interaction
\end{abstract}

%\keywords{}

\maketitle

\section{Introduction}

If there is a direct interaction between two systems, classical light can generate entanglement between them \cite{PhysRevLett.74.4091,PhysRevLett.81.3631}.
%(CITE e.g. ion-trap or Rydberg atoms in general. FIND more works on it). 
%(CITING a general statement or proof on it may be better)
However, when the two parties are not coupled, they cannot be entangled by classical light, 
which only allows a local unitary transformation on each party. 
However, quantum light can be used to create entanglement between two noninteracting systems \cite{Orszag:10}.
%Even if the two parties cannot communicate directly, they can talk to quantum light and retrieve information on the other party by observing the light.
%Matter can exchange quanta with quantum light.
%Apart from the single-photon state, 
Several types of quantum light have been considered to generate entanglement between qubits.
%There are several types of quantum light. 
%One can construct a specific quantum light source for a desired task. 
%However, one can also utilize an ubiquitous resource, the quantum vacuum. 
%Matter can communicate with quantum light, and by using a spatial confinement, a cavity, the communication can become more effective \cite{haroche1989cavity}. 
Especially, the generation of entanglement between two noninteracting two-level systems (or qubits) based on cavity electrodynamics has been studied for various states of quantum light, including the Fock \cite{Cai_2006}, thermal \cite{PhysRevA.65.040101}, coherent \cite{Jian_2004,Jarvis_2009}, and squeezed state \cite{Zou_2005,Orszag:10}. 
%For all of these works, the state of the cavity mode is assumed to be given without specifying how to prepare the state, which is essential for an experimental verification.
% [TODO] This statement should be confirmed by reading each paper throughly.

As one of the most effective and simple methods, a single-photon state of the cavity mode can be used to entangle two qubits. This can be done by exchanging a quantum, which is in this case a photon, between the cavity mode and the qubits. 
%There is only a single quantum in the system but there are two qubits to which the quantum can be transferred. 
%Either qubit can receive the photon, but not both qubit
Suppose both qubits are in their ground states.
Since there is only one photon for two qubits, only one qubit or the other, but not both at the same time, can receive the photon to get excited.
Thus, the resulting state is the superposition of those two possibilities, which is an entangled state between the qubits.
%The entanglement between two systems can be established by exchanging quanta between 
Pumping only a single quantum in a cavity typically requires an external single-photon source \cite{PhysRevLett.78.3221,RevModPhys.87.1379}, which is a quantum state with no classical counterpart.
%in the cavity typically requires an external single-photon source 
%[CITE. a paper on the interface between the cavity mode and qubits]"
%(maximal entangelmetn can be achieved by a single-photon state, which is a highly quantum states (CITE why is this a highly quantum state, possibly the g2 or Glauber) - then, )
%(pose the question, how much entanglement can be obtained by a classical light source?)
%With a single-photon state in a cavity mode coupled to the two qubits, maximally entangled
%Maximally entangled state can be obtained if 
With such a quantum light source, even a maximally entangled state can be achieved.

Then, how much entanglement can be generated if we use a classical light source for pumping the cavity mode? 
%Would it be possible to still achieve the maximal two-qubit entanglement?
This is the question to investigate in this paper.
%In this paper we investigate this question for a pulsed classical light.
%two pulse durations 
We consider an exactly solvable model of two qubits interacting with a single cavity mode \cite{PhysRev.170.379} and compute the entire time-dependent state. 
We explore the resulting two-qubit concurrence \cite{PhysRevLett.78.5022,PhysRevLett.80.2245} as a measure for their entanglement. Previously, this has been done for several initial states of the system \cite{Cai_2006,PhysRevA.65.040101,Jian_2004,Jarvis_2009,Zou_2005,Orszag:10}.
In this work, we do not assume a particular initial state other than the ground state of the total system. Instead, we drive the cavity mode with an external classical field and investigate what kind of state can be prepared.
Since the cavity mode is driven by a classical light, a coherent state may be expected to a good approximation if the interaction of the cavity mode with the qubits is negligible. Some correction may be needed since the cavity mode would interact with the qubits even during the driving.
%We prepare the initial state of the cavity mode by adding a classical external driving field to the system. 
%We thus control the entanglement dynamics of two qubits with classical light source.
We show analytically how the joint state of the cavity mode and the two qubits depends on the external classical field. The interaction between the cavity mode and the qubits is considered consistently with the external driving of the cavity mode, to identify the classes of states that can be prepared with a classical light source.
We study then the entanglement generated by the prepared state.
We demonstrate how the two-qubit entanglement dynamics can be controlled in terms of the strength, duration, phase and temporal shape of the classical light field.
%prepared by the classical light source.

There are two regimes that can be distinguished in terms of the duration of the driving. 
Namely, it can be short or long compared to a characteristic timescale which we denote as $T_{g}$.
%We first discuss the characteristic timescale of the cavity-qubits system, denoted $T_{g}$, which
The timescale $T_{g}$ determines how fast the entanglement is generated after the system is pumped. $T_{g}$ is determined by how strong the cavity mode and each qubit are coupled. When there is no coupling, no quantum can be exchanged and thus no entanglement is generated, which means $T_{g} \rightarrow \infty$. When there is a coupling, quanta can be exchanged and thus entanglement can be generated. The stronger the coupling is, the faster the exchange of quanta would be, which means $T_{g}$ gets shorter. Precise expression of $T_{g}$ is discussed.
%The timescale $T_{g}$ is inversely proportional to the coupl
%
%The timescale $T_{g}$ defines the two regime. One is short and the other is long compared to $T_{g}$. 
%
%We consider an impulsive driving using a subcycle pulse, which is defined by its duration being shorter than the characteristic timescale of the cavity-qubit interaction.
%whose duration is shorter than the characteristic timescale 
%$g^{-1}$ 
%, where the $g$ is the coupling between each qubit and the cavity mode.
%The subcycle pulse is defined 
%Being subcycle is defined to be 
%We derived analytical expressions of the time-evolution of the total system for this highly diabatic driving. 
Once the characteristic timescale $T_{g}$ is identified, we investigate the two mentioned driving regimes.
%In the former regime, 
%the cavity mode is driven by a pulse which is sufficiently short compared to $T_{g}$. 
In the regime when the cavity mode is driven by a pulse which is sufficiently short compared to $T_{g}$, we study how the two-qubit entanglement depends on the pulse strength, duration and the shape. 
%We show that the pulse shape is crucial in determining the entanglement dynamics
We show that by selecting an appropriate pulse shape, the pumping can result in a displacement of the cavity mode or a rotation of the qubits, to a good approximation. The entanglement dynamics can be controlled by selecting the type of pumping through the pulse shape.
In the latter regime, one can adiabatically generate a squeezed state and rotated qubits. The effect of the squeezing and the rotation on the entanglement formation is investigated.

This paper is organized as follows. In Sec. \ref{sec:model}, we describe the model system, where a cavity mode is driven by a classical light source. In Sec. \ref{sec:exchange-quanta}, we show how the entanglement can be generated by exchanging quanta between the qubits and the cavity mode. The characteristic timescale $T_{g}$ of the cavity-qubit interaction is identified. In Sec. \ref{sec:entanglement-by-subcycle-driving}, we consider one of the regimes where the driving duration is sufficiently short with respect to the characteristic timescale of the cavity-qubit interaction. In Sec. \ref{sec:entanglement-generation-by-quasistatic-driving}, we investigate the other regime where the driving is quasistatic. In Sec. \ref{sec:discussion} and \ref{sec:conclusion} we discuss a set of parameters for an experimental realization and conclude the paper with a summary.

\section{\label{sec:model}Model}
We consider two qubits coupled to a resonant cavity mode of frequency $\omega$. A classical external light drives the cavity mode. 
In a rotating frame at the frequency $\omega$, the model Hamiltonian can be written as
\begin{equation}\label{eq:H}
	H(t) = H_{g} + H_{e}(t).
\end{equation}
The first term, 
%of Eq. (\ref{eq:H}), 
\begin{equation}\label{eq:Hg}
	H_{g} = \hbar g (\sigma^{+}a + \sigma^{-}a^{\dagger}),
\end{equation}
describes the interaction between the cavity mode and each qubit under the rotating wave approximation (RWA), which is justified for $g \ll \omega$.
The two-qubit Pauli operator is defined as
\begin{equation}\label{eq:sigma-pm-ladder}
	\sigma^{\pm} = \sigma_{A}^{\pm} + \sigma_{B}^{\pm},
\end{equation}
where $\sigma_{A}^{\pm}$ and $\sigma_{B}^{\pm}$ are the ladder operators for qubits $A$ and $B$, respectively.
$a$ and $a^{\dagger}$ are the annihilation and creation operator of the cavity photon, respectively. 
The second term of Eq. (\ref{eq:H}) is given as
\begin{equation}\label{eq:He_rotframe}
	H_{e}(t) = \hbar\Omega f(t) x_{\omega}(t),
\end{equation}
where $\Omega$ is the driving strength and $f(t)$ is the temporal shape of the external field. $x_{\omega}(t)$ is the quadrature operator defined as
\begin{equation}\label{eq:x_omega_t}
	x_{\omega}(t) \equiv a e^{-i\omega t} + a^{\dagger} e^{i\omega t},
\end{equation}
corresponding to the (normalized) electric field of the cavity mode. The interaction Hamiltonian $H_{e}(t)$ describes a linearly driven oscillator, representing a cavity mode coupled to an external field. The Hamiltonian has been used theoretically \cite{Alsing_1991,PhysRevA.45.5135,PhysRevLett.100.014101,PhysRevA.102.033729} and demonstrated experimentally \cite{Feng2015}.
We consider a pulsed driving. Let $\tau_{d}$ be the duration of the pulse and $t=0$ be the center of the pulse. 
%For a given temporal width $T > 0$, we let $t=-T$ and $t=T$ be the initial and the final time point of the description, respectively. 
The total considered time interval shall be $[-T,T]$, where
\begin{equation}\label{eq:Tu}
	T / \tau_{d} \equiv T_{u} \gg 1,
\end{equation}
%$T = T_{u} \tau_{d}$ 
%\begin{subequations}
%\begin{align*}
%	\label{eq:Tu}
%	T & = T_{u} \tau_{d},\\
%	T_{u} & \gg 1,
%\end{align*}
%\end{subequations}
%for $T_{u} \gg 1$ 
in order to make this time interval long enough to accommodate the pulse.

Consider a pulse with a central frequency $\omega$ which is resonant to the cavity mode and each qubit. Let $f_{0}(t)$ be the envelope of the pulse shape and $\phi$ be the carrier-envelope offset phase. We write the pulse shape as
\begin{equation}\label{eq:ft}
	f(t) = f_{0}(t) \cos(\omega t + \phi).
\end{equation}
%$\tau_{d}$ is the duration of the envelope shape $f_{0}(t)$.
The envelope function can be expanded in a complete set of localized functions, e.g. a set of Hermite-Gaussian (HG) functions,
%Let $f_{\mathrm{HG},m}(t)$ be the $m$-th order HG function
which can be written as
\begin{equation}\label{eq:hg_m}
	f_{\mathrm{HG},m}(u) = N_{m} H_{m}(u) e^{-u^{2}/2}.
\end{equation}
Here, $u \equiv t / \tau_{d}$ and $H_{m}(u)$ is the $m$-th order Hermite polynomial for $m \ge 0$. The normalization factor $N_{m}$ is given by
\begin{equation*}
	N_{m} = \frac{\pi^{-1/4}}{\sqrt{m!\,2^{m}}},
\end{equation*}
so that
\begin{equation*}
	\int_{-\infty}^{\infty}|f_{\mathrm{HG},m}(u)|^{2} = 1.
\end{equation*}

\section{Exchanging quanta generates entanglement}\label{sec:exchange-quanta}

How can the two qubits become entangled?
One way is to exchange a quantum with the cavity mode.
%Once we have a set of quanta to exchange, 
%The entanglement between qubits can be generated by exchanging 
For example, suppose there is no driving and consider an initial state where all the qubits are in their ground state and the cavity mode has one photon. In this case, there is only one excitation in the system, a photon. Due to the coupling between each qubit and the cavity mode, Eq. (\ref{eq:Hg}), the quantum starts to `move' from the cavity mode to the qubits. However, there are two qubits for a single quantum. Since the coupling strength between each qubit and the cavity mode is the same, the probability that the quantum will be found after some time at one of the qubits is identical as for the other qubit. This state, where the two possibilities of a bipartite system (two qubits) are superposed, is entangled.

We shall trace this entanglement generation. Let us denote the initial state $|\psi(0)\rangle$ as $|00;1\rangle \equiv |00\rangle|1\rangle$. $|00\rangle \in \mathcal{H}_{q}$ represents the state of two qubits where both of them are in their ground states. $|n\rangle$ with $n \ge 0$ denotes the Fock state with $n$ photons in the cavity mode. $\mathcal{H}_{q}$ and $\mathcal{H}_{\gamma}$ represent the Hilbert space of the qubits and of the cavity mode, respectively. At time $t$, the state evolves into a certain state, denoted $|\psi(t)\rangle$. The time evolution is governed by the Hamiltonian $H_{g}$ in Eq. (\ref{eq:Hg}). One can diagonalize the Hamiltonian to calculate the exact expression of $|\psi(t)\rangle$. However, we note that $|\psi(t)\rangle$ would be a superposition of only two states, $|00;1\rangle$ and $|\Psi^{+};0\rangle \equiv |\Psi^{+}\rangle|0\rangle$, where $|\Psi^{+}\rangle \equiv (1/\sqrt{2})(|01\rangle + |10\rangle)$. This can be seen by noting that $\sigma^{+} | 00\rangle = \sqrt{2} | \Psi^{+}\rangle$ and that $H_{g}$ consists of two terms, $\sigma^{+}a$ and $\sigma^{-}a^{\dagger}$, which describe exchanges of quanta between both qubits and the cavity mode. Considering a time evolution with a finite time $t$ as a succession of infinitesimal steps $\Delta t$, each approximated as $1 +(-i/\hbar)H_{g}\Delta t$, all possible paths that a state may evolve along can be indicated by the following diagram:
\begin{equation}\label{eq:diagram-single-quantum}
	0
	\xleftharpoondown[\sigma^{-}a^{\dagger}]{} 
	|00\rangle |1\rangle
	\xrightleftharpoons[\sigma^{-}a^{\dagger}]{\sigma^{+}a}
	|\Psi^{+}\rangle |0\rangle
	\xrightharpoonup[]{\sigma^{+}a}
	0.
\end{equation}
From this diagram, one can expect that the state will be a superposition of the two states.
% $|00;1\rangle$ and $|\Psi^{+};0\rangle$ in general. 
An exact calculation shows that
\begin{equation*}
	U_{g}(t) |00;1\rangle
	= \cos{(g_{1}t)} |00;1\rangle + \sin{(g_{1}t)} | \Psi^{+}; 0 \rangle,
%	\cos
%	|\psi(t)\rangle = 
\end{equation*}
where $U_{g}(t) = \exp\left[ -\frac{i}{\hbar}H_{g}t \right]$ and $g_{1} = \sqrt{2}g$. 
%For the derivation, see Appendix \ref{sec:Ug-and-fock-state-evolution}.
At $t=0$, the total state is $|00;1\rangle$ and the two qubits are not entangled. When $t = \pi / 2g_{1}$, the total state becomes $|\Psi^{+};0\rangle$, where the two qubits are in a maximally entangled state.
Note that the timescale of the entanglement dynamics is proportional to $g_{1}^{-1} \sim g^{-1}$.
%(describe the case of a single-photon and ground-qubit state as the initial state)

When there are $n \ge 2$ quanta, the state can have an additional component, namely $|11;n-2\rangle \equiv |11\rangle |n-2\rangle$, which can be noticed by considering the possible paths as the following diagram:
\begin{equation*}
	0
	\xleftharpoondown[\sigma^{-}a^{\dagger}]{} 
	\!|00\rangle |n\rangle
	\xrightleftharpoons[\sigma^{-}a^{\dagger}]{\sigma^{+}a}
	|\Psi^{+}\rangle |n-1\rangle
	\xrightleftharpoons[\sigma^{-}a^{\dagger}]{\sigma^{+}a}
	|11\rangle |n-2\rangle
	\xrightharpoonup[]{\sigma^{+}a}
	0.
\end{equation*}
An exact calculation shows that the dynamics timescale is proportional to $g_{n}^{-1} \sim (\sqrt{n} g)^{-1}$, where 
\begin{equation}\label{eq:g_n}
	g_{n} = \sqrt{4n-2}\,g
\end{equation}
for $n \ge 1$. Here, we define a timescale, denoted as $T_{g}$, of the system containing $n$ quanta as
\begin{equation*}
	T_g = (\sqrt{n}g)^{-1}.
\end{equation*}
The dynamics of observables for a state with $n$ quanta will characteristically unfold at this timescale. We expect that the formation of entanglement takes about this amount of time when there are $n$ quanta in the system.

To confirm the timescale of the entanglement dynamics, we quantify the entanglement of the reduced density operator $\rho$ of the two qubits. The density operator is defined as
\begin{equation}\label{eq:rho}
	\rho(t) = \mathrm{tr}_{\gamma}[|\psi(t)\rangle\langle \psi(t)|],
\end{equation}
where $|\psi(t)\rangle$ represents the state of the total system at time $t$ and $\mathrm{tr}_{\gamma}$ is a partial trace with respect to the degree of freedom of the cavity mode.
After tracing out the cavity mode, the qubits are in a mixed state in general. A mixed state can be represented as a statistical ensemble of pure states with their associated probabilities. 
Each possible pure state in the ensemble has a well-defined entanglement, defined via the von Neumann entropy that represents the upper bound on the purification/entanglement cost, the latter being defined in terms of the cooperative game that uses the Local Operations and Classical Communication (LOCC) protocols \cite{PhysRevA.53.2046}. It has been also demonstrated \cite{PhysRevA.53.2046} that the von Neumann entropy can only go down in average when non-unitary operations, such as measurements, are performed.
%Each possible pure state in the ensemble has a well-defined entanglement. Using the probability distribution and the entanglement of each pure state, the average value of the entanglement can be calculated. However, the number of ensembles corresponding to the given mixed state density matrix is not unique. Depending on the choice of the ensemble, the average entanglement varies as well.

%The entanglement of formation is defined as the minimal average entanglement out of all possible statistical ensembles which correspond to the given mixed-state density matrix. This quantity is defined for a general system and it is not trivial to find its value. However, the entanglement of formation of two qubits, is known to be related to a quantity called `concurrence', which can be calculated explicitly for a given density matrix \cite{PhysRevLett.80.2245}. Zero concurrence means no entanglement of formation and concurrence of 1 implies that the system is in a maximally entangled state. 

To quantify entanglement of mixed states of two qubits, the notion of entanglement of formation has been introduced 
%to mixed states of two qubits \cite{PhysRevA.54.3824}. It was formally defined 
as the minimal average von Neumann entropy of an ensemble of pure states that represents the given mixed state (the pure states do not have to be mutually orthogonal, and their number is not fixed), and the minimum is taken over all such ensembles \cite{PhysRevA.54.3824}. It has been shown in Ref. \cite{PhysRevA.54.3824} that the entanglement of formation defined in this way has the analogous property of decreasing upon non-unitary transformations, associated with measurements, and therefore is considered as a good measure of entanglement for mixed states of two qubits.

Since the definition involves an optimization problem, the entanglement of formation is apparently hard to compute. Therefore, an important result is an explicit formula for the entanglement of formation in terms of the so-called concurrence, postulated in \cite{PhysRevLett.78.5022}. There, the concurrence was defined in terms of the eigenstates of a matrix acting in the Hilbert space of two qubits. This matrix is composed of the product of the density matrix of the given mixed state and its involuted counterpart, with the involution that comes from the anti-linear operator acting in the Hilbert space of a single qubit that represents the time-reversal symmetry. The formula for the entanglement of formation in terms of the concurrence has been proven there, for a particular case of the density matrices with at least two zero eigenvalues. The proof has been extended to a general mixed state of two qubits in \cite{PhysRevLett.80.2245}.

The calculation of concurrence is related to a time-reversal operation. For qubits, which are pseudospins, this corresponds to a `spin-flip'. The concurrence is defined in terms of how similar is a state to its time-reversed, or spin-flipped counterpart. For example, $|00\rangle\langle00|$ is a product state.
Its spin-flipped counterpart is $|11\rangle\langle11|$.
%$(\sigma_{A}^{y}\sigma_{B}^{y})|00\rangle = -|11\rangle$. 
The similarity between the two states is quantified by the absolute value of their inner product, namely, $|\langle 11 | 00\rangle| = 0$, which is consistent with the zero entanglement of the state $|00\rangle$. If one does the same procedure for a maximally entangled state, say, $|\Psi^{+}\rangle\langle\Psi^{+}|$, one notices that its spin-flipped counterpart is the same, thus yielding the maximal similarity $|\langle\Psi^{+}|\Psi^{+}\rangle| = 1$.
For a mixed state, the concurrence is defined as
\begin{equation}\label{eq:concurrence}
	C \equiv \max\{0, \tilde{C}\},
\end{equation}
which is either $0$ or a quantity called `naive' concurrence $\tilde{C}$. The naive concurrence is given as
\begin{equation}\label{eq:naive-concurrence}
	\tilde{C} \equiv \lambda_{1} - \lambda_{2} - \lambda_{3} - \lambda_{4},
\end{equation}
where $\lambda_{1} \ge \lambda_{2} \ge \lambda_{3} \ge \lambda_{4}$ are square roots of the eigenvalues of $\rho\tilde{\rho}$. Here, $\tilde{\rho} \equiv \sigma_{A}^{y}\sigma_{B}^{y} \rho^{*} \sigma_{A}^{y}\sigma_{B}^{y}$ is a spin-flipped counterpart of $\rho$. The multiplication of $\rho$ with $\tilde{\rho}$ and calculating its eigenvalues quantifies the similarity between $\rho$ and $\tilde{\rho}$. By combining with signs in a special way as given in Eq. (\ref{eq:naive-concurrence}), it is known that the concurrence defined as Eq. (\ref{eq:concurrence}) indeed is a measure of the entanglement of formation \cite{PhysRevLett.80.2245}.

We denote the concurrence as $C_{n} \equiv C_{n}(t)$ for the state $|\psi_{n}(t)\rangle = U_{g}(t)|00;n\rangle$ with $n \ge 0$. We get
\begin{equation}\label{eq:concurr-n}
	C_{n} =
	\begin{cases}
			0 & (n=0)\\
			\max\left\{0, \rho_{n}^{\Psi^{+}} - 2\sqrt{\rho_{n}^{00} \rho_{n}^{11}}\right\} & (n\ge1)
	\end{cases},
%	C_{n}(t) = 
%	\begin{cases}
%		1 & (n=0)\\
%		\max\{ 0, \rho_{\Psi^{+}}(t) - 2\sqrt{\rho_{00}(t)\rho_{11}(t)} \} & (n \ge 1)
%	\end{cases},
\end{equation}
where $\rho_{n}^{\mu} \equiv \rho_{n}^{\mu}(t) \equiv \langle \mu | \rho_{n}(t) | \mu \rangle$ is the population of a two-qubit state $|\mu\rangle$ for $\mu \in \{00, \Psi^{+}, 11\}$. $\rho_{n}(t)$ is the reduced density operator of the qubits, which is defined as Eq. (\ref{eq:rho}) with $|\psi(t)\rangle = |\psi_{n}(t)\rangle$. The populations are given as
\begin{equation}\label{eq:rho-populations}
	\begin{split}
		\rho_{n}^{00}(t) & = [p_{n} + q_{n} \cos(g_{n}t)]^{2}\\
		\rho_{n}^{\Psi^{+}}(t) & = q_{n} \sin^{2}{(g_{n}t)}\\
		\rho_{n}^{11}(t) & = p_{n}q_{n} [ 1 - \cos(g_{n}t) ]^{2},
	\end{split}
\end{equation}
where $p_{n} = (n-1)/(2n-1)$, $q_{n} = n / (2n-1)$ and $g_{n}$ is defined by Eq. (\ref{eq:g_n}).

\begin{figure}
	\centering
	\includegraphics[width=1\linewidth]{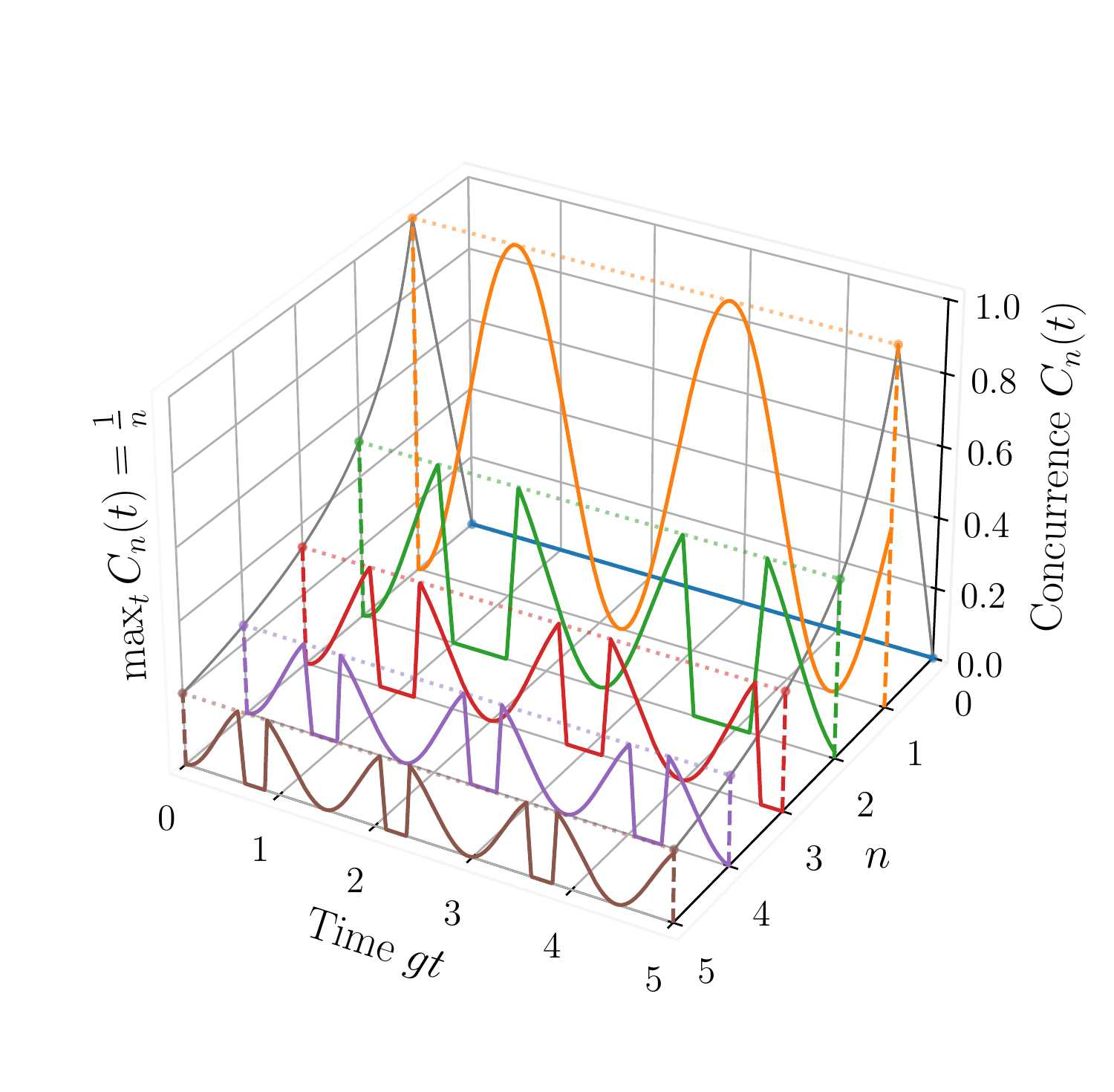}
	\caption{The concurrence $C_{n}(t)$, as given by Eq. (\ref{eq:concurr-n}), when there are $n$ photons in the initial state. The gray solid lines shown on the photon-number--concurrence planes represent the maximal concurrence, as given by Eq. (\ref{eq:maximum-values}).}
	\label{fig:concurrences-test-023}
\end{figure}

The maximal concurrence is achieved when there is only $n=1$ photon in the initial state. To see this, 
we notice that $C_{1}(t) = \sin^{2}(g_{1}t)$, which follows from Eqs. (\ref{eq:concurr-n}) and (\ref{eq:rho-populations}). Similarly, the maximal value achievable for each $n$ can be derived as,
\begin{equation}\label{eq:maximum-values}
	%	C_{n}^{\mathrm{max}} \equiv 
	\max_{t}C_{n}(t) =
	\begin{cases}
		0 & (n=0)\\
		1/n & (n\ge1)
	\end{cases},
\end{equation}
which shows that the entanglement vanishes as $n \rightarrow \infty$.
In Fig. \ref{fig:concurrences-test-023}, we plot the concurrence $C_{n}(t)$ and its maximal value $\max_{t}{C_{n}(t)}$ for each initial photon number $n$. Although the photon number changes with time, the total number of quanta is conserved, which is a sum of the number of photons and the number of excited qubits. In other words, the state of the system always belongs to a subspace with $n$ quanta. Thus, each concurrence has a well-defined period $2\pi / g_{n} \sim (\sqrt{n}g)^{-1} = T_{g}$, which determines the timescale of entanglement generation in the subspace of $n$ quanta.
%Starting from zero, the concurrence grows at a timescale of 
%Each concurrence results from a state which belongs to a subspace with $n$ quanta.

Regarding the entanglement generation mechanism, we note that what creates or eliminates the entanglement is the set of two ladder operators, $\sigma^{+}$ and $\sigma^{-}$.
%, generating a Bell state $|\Psi^{+}\rangle$ from $|00\rangle$ and $|11\rangle$, respectively. 
%For example, applying $\sigma^{+}$ 
%$\sigma^{+} |00\rangle = |10\rangle + |01\rangle \equiv \sqrt{2} |\Psi\rangle$.
%Especially, the `$+$' sign that connects $\sigma^{\pm}_{A}$ and $\sigma^{\pm}_{B}$ in the expression of $\sigma^{\pm}$, as given in Eq. (\ref{eq:sigma-pm-ladder}), does the role. 
On top of that, what triggers the action of these ladder operators is an event of exchanging a quantum with the cavity mode, see Eq. (\ref{eq:diagram-single-quantum}). Thus, when there is no quantum in the system, such as $|00\rangle |0\rangle$, no exchange of quanta can occur and no entanglement is generated.

\section{
	Entanglement generation by subcycle driving, $\tau_{d} \ll T_{g}$
}\label{sec:entanglement-by-subcycle-driving}

In this section, we consider entanglement generation by a short pulse. 
With $\tau_{d} \ll T_{g}$, the action of the pulse on the system lasts shorter than 
%the pulse then acts on the system fast enough than 
the timescale of the interaction between the cavity and the qubits.
However, we need to set also a lower bound on $\tau_{d}$.
%In addition, we consider a lower bound of $\tau_{d}$. 
Firstly, there seems to exist a finite lower bound for the pulse duration that can be realized experimentally. Secondly, the pulse shall possess a well-defined carrier frequency $\omega$. This is to prevent exciting other cavity modes and efficiently couple the external mode to the cavity mode with the desired frequency. The condition reads $\Delta\omega \ll \omega$, where $\Delta \omega$ is the bandwidth of the pulse. 
Combining this condition with the uncertainty relation between time and frequency, namely, $(\Delta \omega)^{-1} < \tau_{d}$ (or $\sim \tau_{d}$ for a Fourier-limited pulse), we get $\omega^{-1} \ll \tau_{d}$.
%\begin{equation*}
%	\omega^{-1} \ll \tau_{d}.
%\end{equation*}
%Since the time and the frequency are conjugate each other, 
%the lower bound of the pulse duration is given as
%Firstly, experimental
%Also, there exists a lower bound of $\tau_{d}$.
%One may 
Using both limits, one arrives at the range of pulse durations
\begin{equation}\label{eq:pulse-duration-condition}
	g / \omega \ll g\tau_{d} \ll 1 / \sqrt{n}.
%	\omega^{-1} \ll \tau_{d} \ll (\sqrt{n}g)^{-1}.
\end{equation}
Again, $\sqrt{n}$ is determined by the number of quanta involved in the dynamics of the state, as in Sec. \ref{sec:exchange-quanta}.

%(write the regime of the pulse duration: $\omega^{-1} \ll \tau_{d} \ll g^{-1}$).

In this short-pulse regime, the interaction between cavity mode and the qubits would seem almost frozen during the pulse. Since only the cavity mode is externally driven, the qubits can notice the effect of the pulse only through the state of the cavity mode, which is coupled to the qubits. The higher the cavity mode-qubit coupling $g$, the faster can the changes in the cavity mode affect the qubits. As described in Sec. \ref{sec:exchange-quanta}, the interaction speed is roughly proportional to $\sqrt{n}$, where $n$ is on the same magnitude as the number of quanta in the state undergoing the dynamics. Thus, if the pulse duration $\tau_{d}$ is sufficiently shorter than the interaction timescale $T_{g} = (\sqrt{n}g)^{-1}$, then in the leading order we can leave the qubits out of consideration while the cavity is pumped.
%the effect of the qubits interacting with the cavity mode can be neglected for a leading-order approximation.
%barely notices the driving pulse if the pulse duration is 
%The cavity mode would be driven by the pulse
%the interaction between the cavity mode and the qubits can be neglected to obtain the leading-order effect.
%the state of the cavity mode changes due to the driving pulse pulse. Since 

Formally, neglecting the cavity mode-qubit interaction translates into $g \rightarrow 0$.
%dropping the Hamiltonian $H_{g}$ given as Eq. (\ref{eq:Hg}), which describes the interaction between each qubit and the cavity mode. 
In this case, the total Hamiltonian in Eq. (\ref{eq:H}) reduces to $H(t) = H_{e}(t)$, which is a linearly driven harmonic oscillator in the rotating frame. Classically, it has an exact solution obtained by solving the Hamilton equations. In the original frame it reads
\begin{equation}\label{eq:hamilton-x-and-p}
\begin{split}
	\dot{x}(t) & = +\omega p(t),\\
	\dot{p}(t) & = -\omega x(t) - 2\Omega f(t),
\end{split}
\end{equation}
with the following classical-quantum correspondence, $x \leftrightarrow a + a^{\dagger}$, $p \leftrightarrow -ia + ia^{\dagger}$. Expressing the two real variables $x$ and $p$ by a single complex variable $z_{\omega} \equiv (x + ip) / 2$, the Hamilton equations Eq. (\ref{eq:hamilton-x-and-p}) can be combined as
\begin{equation}\label{eq:hamilton-z}
	\dot{z}_{\omega}(t) = -i \omega z_{\omega}(t) -i \Omega f(t).
\end{equation}
In the absence of driving, i.e. $\Omega = 0$, the system exhibits a harmonic motion, $z_{\omega}(t) = e^{-i\omega (t-t_{0})} z_{\omega}(t_{0})$, for a given reference time point $t_{0}$. In the rotating frame defined by $z_{\omega}(t) \equiv e^{-i{\omega}t} z(t)$, the Hamilton equation for $z_{\omega}(t)$ given by Eq. (\ref{eq:hamilton-z}) translates to
\begin{equation*}
	\dot{z}(t) = -i \Omega f(t) e^{i\omega t}.
\end{equation*}
The solution describes a displacement with an amplitude,
\begin{equation}\label{eq:z-t-classical}
	z(t) - z(t_{0}) = -i \Omega \int_{t_{0}}^{t} f(t') e^{i\omega t'}.
\end{equation}
Although this result comes from the classical Hamilton equations, Eq. (\ref{eq:hamilton-x-and-p}), the same expression can be obtained from a fully quantum-mechanical description, 
governed by $H(t)$
%by solving for $H(t)$ 
in Eq. (\ref{eq:H}) with $g \rightarrow 0$. Even if $g$ is not exactly zero, the displacement with the amplitude given in Eq. (\ref{eq:z-t-classical}) is a good approximation as long as the pulse is sufficiently shorter than $(\sqrt{n}g)^{-1}$. 
In this approximation, the relevant number of quanta $n$ at time $t$ is around the average photon number $|z(t)|^{2}$. Thus, self-consistency requires
\begin{equation*}
	\tau_{d} \ll (\sqrt{n}g)^{-1} \sim (|z(t)| g)^{-1}
\end{equation*}
for all relevant times $t$. 
%Equivalently, we set $|z| \ll (g\tau_{d})^{-1}$, meaning that 
We may increase the amplitude $z$ until
%by using a short enough pulse that satisfies 
$g\tau_{d} \ll 1/|z|$ still holds. In order to generate a large amplitude with a certain level of accuracy, a sufficiently short pulse is required. Exactly how small the duration should be is determined by the pulse shape. If we can understand how the accuracy depends on the pulse shape, we may be able to find a pulse shape which generates a large enough amplitude with a good fidelity, even for a moderately short pulse. In Sec. \ref{sec:towards-exact-operations}, we express the fidelity as a functional of the pulse shape and utilize it to properly tailor the pulse for mitigating the error. We expect the error to come from the fact that we neglected the interaction between the qubits and the cavity. This is shown in Sec. \ref{sec:second-leading-order-effect-rotation}. The magnitudes of this second-order and higher-order terms are identified in the following sections. Generally, these terms occur to be smaller than the leading-order term. However, they can become essential when $z(t)$ converges to zero after the end of the pulse so that the leading-order term vanishes. We discuss the condition for such pulses with almost no displacement and generalize the idea so that one can switch on or off a term with a specific order. This can be useful since each term has its own signature. For example, the leading-order term induces a displacement of the cavity mode and the second-order term induces a rotation of the qubits.
%These terms must be included when the $z(t)$ converges to zero after the end of the pulse.
%This interaction term $H_{g}$ becomes dominant when the $z(t)$ converges to zero after the end of the pulse.
%so that one can realize the displacement operations with 
%One may enhance this bound a pulse shaping

%(mention, at the end of the derivation for the amplitude z, be sure to be self-consistent in that the initial assumption, $\tau_{d} \ll T_{g} = (\sqrt{n}g)^{-1}$ is fulfilled by the final result, z, whose average photon number is given as $|z|^{2}$ so that the $\sqrt{n} \sim \sqrt{|z|^{2}} = |z|$. It implies this argument can be self-consistent for a weak field satisfying $\tau_{d}T_{g} \sim \tau_{d}|z|g \ll 1$. Equivalently, we get $|z| \ll (g\tau_{d})^{-1}$, meaning that we may increase the amplitude $z$ by using a short enough pulse duration, which satisfies $g\tau_{d} \ll 1/|z|$. One may enhance (relax) this bound a pulse shaping).

%when the external driving field has a short duration compared to the timescale of the entanglement is built up.
In this section, we show that the leading-order effect of the pulse is to create a coherent state in the cavity mode. Further, we formulate the conditions when this effect can be turned off by pulse shaping. Then, the second-order term becomes relevant. It acts only on the state of the qubits. We demonstrate that by shaping the pulse appropriately, one can select which part of the system is pumped, either the cavity mode or the qubits. 
%This relation between the pulse shape 
Understanding how the pulse shape controls both the amplitude of the generated coherent state and the states of the qubits is essential for an experimental realization of the entanglement generation as well as other phenomena including the collapse and revival of the qubits observables \cite{PhysRevLett.44.1323} and the existence of the `attractor' state \cite{Jarvis_2009,PhysRevLett.65.3385}.
%the insentitivity of the qubits state to the initial qubit state.
%either the cavity mode or the qubits, which is essential for experiments.

%\subsection{title}

%(NOTE: In Fig. \ref{fig:naive-concurr-test-90}, add the a set of insets showing the state right before the pulse, right after the pulse and at some special time point such as half of the revival time and the revival time. Show the Bloch sphere and the Wigner quasi distribution function if the state is separable at the time points, or can be represented with upto few terms if entangled.)
%(NOTE, also, add a magnified f(t) as another inset)

%\begin{figure}
%	\centering
%	\includegraphics[width=1\linewidth]{fig/naive-concurr-test-90}
%	\caption{Naive concurrence induced by an ultrashort pulse.}
%	\label{fig:naive-concurr-test-90}
%\end{figure}

\subsection{Interaction picture for the pulse}

%We start from deriving the leading-order term. 
In order to describe the effect of the pulse, we switch to a variant of the interaction picture. This is done as follows: in the absence of external driving, i.e. $\Omega = 0$, the Hamiltonian, Eq. (\ref{eq:H}), becomes time-independent, $H(t) = H_{g}$.
The time evolution from the initial time $-T$ to $t \in [-T,T]$ can be described by the total time-evolution operator
\begin{equation*}
	U(t,-T) = U_{g}(t,-T) = U_{g}(t,0) U_{g}(0,-T),
\end{equation*}
where $U(t,-T)$ and $U_{g}(t,t') \equiv U_{g}(t-t') \equiv \exp[-i H_{g} (t-t') / \hbar]$ are the time-evolution operators generated by $H(t)$ and $H_{g}$ respectively. In the presence of an external pulse, i.e. $\Omega > 0$, centered at time $t=0$, we define a time-evolution operator $\mathcal{U}$
\begin{equation}\label{eq:U-decomposed}
	U(t,-T) \equiv U_{g}(t,0) \, \mathcal{U}(t; 0,-T) \, U_{g}(0,-T).
\end{equation}
The time-evolution operator $\mathcal{U}$ accounts for the effect of the pulse. For brevity, we denote $\mathcal{U}(t; 0,-T)$ as $ \mathcal{U}(t)$. $\mathcal{U}(t)$ satisfies
\begin{equation}\label{eq:tdse-U-H_I}
	\dot{\mathcal{U}}(t) = -\frac{i}{\hbar}H_{I}(t) \mathcal{U}(t).
\end{equation}
Here $H_{I}(t)$ is the Hamiltonian in the interaction picture:
\begin{equation}\label{eq:H_I_t}
\begin{split}
	H_{I}(t) 
	& = U_{g}^{\dagger}(t) [H(t) - H_{g}] U_{g}(t)\\
%	& = \hbar \Omega f(t) \tilde{H}_{I}(t).
	& = \hbar \Omega \tilde{H}_{I}(t),
\end{split}
\end{equation}
with 
%Eqs. (\ref{eq:H}) and (\ref{eq:He_rotframe}) are used to derive the second line of Eq. (\ref{eq:H_I_t}), where the normalized Hamiltonian is given as
\begin{equation}\label{eq:tilde_H_I_t}
%	\tilde{H}_{I}(t) = U_{g}^{\dagger}(t) \, x_{\omega}(t) \,U_{g}(t).
	\tilde{H}_{I}(t) = f(t) \,U_{g}^{\dagger}(t) \, x_{\omega}(t) \,U_{g}(t).
\end{equation}
Since $\tilde{H}_{I}(t)$ is proportional to the pulse shape $f(t)$, 
it impacts the evolution of the system
%this Hamiltonian is effective 
only for a short duration of time, $\tau_{d}$. If the pulse is sufficiently shorter than the cavity-qubits timescale, i.e. $\tau_{d} \ll T_{g}$, then $U_{g}(t)$ in Eq. (\ref{eq:tilde_H_I_t}) essentially remains identity during the interaction with the pulse, 
%In this case, we may be able to set 
so that $\tilde{H}_{I}(t) \simeq f(t) x_{\omega}(t)$.
%We note that 
Formally, this can be seen by using the identity of Campbell \cite{Campbell}, 
\begin{equation}\label{eq:campbell}
	e^{X}Ye^{-X} = Y + [X,Y] + \frac{1}{2}[X,[X,Y]] + \cdots,
\end{equation}
%$e^{X}Ye^{-X} = Y + [X,Y] + (1/2)[X,[X,Y]] + \cdots$, 
to expand $\tilde{H}_{I}(t)$ in powers of $g\tau_{d}$,
\begin{equation}\label{eq:tilde_H_I_g-tau_d-expansion}
%	\tilde{H}_{I}(u) \equiv \sum_{k=0}^{\infty} (g\tau_{d})^{k} \tilde{H}_{I}^{(k)}(u).
\begin{split}
	\tilde{H}_{I}(u) 
	= &\,(g\tau_{d})^{0} f(u) x_{\omega}(u) \\
	+ &\,(g\tau_{d})^{1} f(u) u [i\tilde{H}_{g}, x_{\omega}(u)] \\
	+ &\,\mathcal{O}[(g\tau_{d})^{2}],
\end{split}
%\tilde{H}_{I}(u) = f(u) x_{\omega}(u) + (g\tau_{d}) f(u) u [i\tilde{H}_{g}, x_{\omega}(u)].
\end{equation}
where $\tilde{H}_{g} \equiv H_{g} / \hbar g$ and $u \equiv t / \tau_{d}$.
%For later purposes, we use the identity of Campbell \cite{Campbell} to expand $\tilde{H}_{I}(t)$ in the power of $g\tau_{d}$, with a change of variable $u \equiv t / \tau_{d}$,
%%we note that $\tilde{H}_{I}(t)$ can be expanded in the power of $g\tau_{d}$ as
%\begin{equation*}
%	\tilde{H}_{I}(u) = \sum_{k=0}^{\infty} (g\tau_{d})^{k} \tilde{H}_{I}^{(k)}(u),
%\end{equation*}
%where 
%\begin{equation*}
%	\tilde{H}_{I}^{(k)}(u) = f(u) \frac{u^{k}}{k!} \mathrm{ad}_{i\tilde{H}_{g}}^{k} x_{\omega}(u),
%\end{equation*}

%\subsection{Leading order}
\subsection{Leading-order effect: displacement of the cavity mode}\label{sec:leading-order-effect-displacement}
The solution of Eq. (\ref{eq:tdse-U-H_I}) can be expressed in terms of the Magnus expansion \cite{Magnus},
\begin{subequations}\label{eq:mathcal-U-magnus}
\begin{align}
	\label{eq:magnus}
	\mathcal{U}(t) & \equiv \exp\left[-i A_{I}(t)\right],\\
	\label{eq:magnus-exponent}
	A_{I}(t) & = \sum_{m=1}^{\infty}A_{I}^{(m)}(t).
	%	\mathcal{U}(t) = \exp[\sum_{n=1}^{\infty}\Omega^{(n)}_{I}(t)].
	%	\mathcal{U}(t) = \exp\left[-i \sum_{n=1}^{\infty}A_{I}(t)\right].
\end{align}
\end{subequations}
The exponent $A_{I}(t)$ of $\mathcal{U}$ is expanded in powers of $\Omega \tau_{d}$ so that
\begin{subequations}\label{eq:A_I_m_t}
\begin{align}
	A_{I}^{(m)}(t) 
	& \equiv (\Omega \tau_{d})^{m} \tilde{A}_{I}^{(m)}(t) \\
	\label{eq:A_I_m_t_magnitude}
	& = \mathcal{O}[(\sqrt{n}\Omega \tau_{d})^{m}]
%	\\
%	& = \mathcal{O}[(\Omega \tau_{d})^{m}],
\end{align}
\end{subequations}
for each $m \ge 1$. For brevity, let us denote $\tilde{\Omega}_{n} \equiv \sqrt{n}\Omega \tau_{d}$. We then get
\begin{equation}\label{eq:A_I_m_t_in_tilde_Omega_n}
	A_{I}^{(m)}(t) = \mathcal{O}[(\tilde{\Omega}_{n})^{m}].
\end{equation}
% A_{I}^{(1)} = \Omega \tau_{d} a + H.c.
% 
The factor $\sqrt{n}$ in Eq. (\ref{eq:A_I_m_t_magnitude}) comes from the fact that the total degree of $\tilde{A}_{I}^{(m)}(t)$ is $m$ in $a$ and $a^{\dagger}$, whose matrix elements in the subspace of $n$ quanta are on the order of $\sqrt{n}$. A derivation of Eq. (\ref{eq:A_I_m_t}) is presented in Appendix \ref{sec:magnus-expansion-order-of-magnitudes}.
%$a$ and $a^{\dagger}$ in $\tilde{A}_{I}^{(m)}(t)$.  $a^{\dagger}$
%Let us denote a normalized term 
% with
%Let us normalize each term by
%\begin{equation}\label{eq:A_I_m}
%	\tilde{A}_{I}^{(m)}(t) \equiv A_{I}^{(m)}(t) / (\Omega\tau_{d})^{m},
%%	A_{I}^{(m)}(t) \equiv (\Omega\tau_{d})^{m} \tilde{A}_{I}^{(m)}(t),
%\end{equation}
%for each $m \ge 1$. 
When $\tilde{\Omega}_{n} \equiv \sqrt{n}\Omega \tau_{d} \ll 1$, the leading-order term is $A_{I}^{(1)}(t)$,
which can be written in terms of
%whose normalized term can be written, with a change of variable $u \equiv t / \tau_{d}$, as
\begin{equation*}
	\tilde{A}_{I}^{(1)}(t) = \int_{-T/\tau_{d}}^{t/\tau_{d}} {du \tilde{H}_{I}(u)}.
\end{equation*}
%In order to see the dependence
%From Eq. (\ref{eq:tilde_H_I_t}), $\tilde{H}_{I}(u')$ depends on $U_{g}(t)$ and $f(t)$. 
In order to see the dominant effect of a short pulse such that $\tau_{d} / T_{g} = \sqrt{n} g\tau_{d} \equiv \tilde{g}_{n}  \ll 1$, we substitute $\tilde{H}_{I}(u')$ with Eq. (\ref{eq:tilde_H_I_g-tau_d-expansion}) and take the leading-order term in $g\tau_{d}$.
%Substituting $\tilde{H}_{I}(u')$ with Eq. (\ref{eq:tilde_H_I_g-tau_d-expansion}), we get the leading-order term
We get then
\begin{equation}\label{eq:A_I_1_t_decomposed}
	\tilde{A}_{I}^{(1)}(t) = \tilde{A}_{I}^{(1,0)}(t) + \mathcal{O}(g\tau_{d}),
\end{equation}
where
\begin{equation}\label{eq:tilde_A_I_1_0_t_explicit}
\begin{split}
	\tilde{A}_{I}^{(1,0)}(t) 
	& \equiv {A}_{I}^{(1,0)}(t) / (\Omega \tau_{d})\\
	& = s_{1}(t)\,a + s_{1}^{*}(t)\,a^{\dagger},
\end{split}
\end{equation}
with
\begin{equation}\label{eq:s_1_t}
	s_{1}(t) = \int_{-T/\tau_{d}}^{t / \tau_{d}}{du f(u) e^{-i \omega \tau_{d} u}}.
\end{equation}
After the pulse, i.e. when $t \gg \tau_{d}$, and with $T$ satisfying Eq. (\ref{eq:Tu}), 
$s_{1}(t)$ becomes
\begin{equation}\label{eq:s1_asympt}
	s_{1} \simeq \hat{f}(\omega \tau_{d})
\end{equation}
where $\hat{f}(k) \equiv \int_{-\infty}^{\infty}{du f(u) e^{-iku}}$ is the Fourier transform of $f(u)$. By controlling the central frequency component of the pulse, one can make $s(t)$ either zero or non-zero for $t \gg \tau_{d}$. 

In our case, the pulse consists of a central frequency $\omega$ and an envelope $f_{0}(t)$ with the carrier-envelope phase $\phi$, as written in Eq. (\ref{eq:ft}). 
Thus, the Fourier component of $f(u)$ at $\omega\tau_{d}$ can be expressed as
%Thus, $s_{1}$ can be expressed as
\begin{equation}\label{eq:f-hat-omega-tau_d}
	\hat{f}(\omega\tau_{d}) 
	= \frac{1}{2}e^{i\phi} \hat{f}_{0}(0) + \frac{1}{2}e^{-i\phi}\hat{f}_{0}(2\omega\tau_{d}),
\end{equation}
where $\hat{f}_{0}(k)$ is the Fourier transform of the envelope function $f_{0}(u)$. Note that $f_{0}(u)$ is defined in the scaled time domain $u \equiv t / \tau_{d}$, with duration 
%so that the width (pulse duration) of $f_{0}(u)$ in the scaled time domain is 
$\tau_{d}/\tau_{d} = 1$. Thus, the width of $\hat{f}_{0}(k)$ in the Fourier domain is also on the order of $1$.
If one considers a pulse with a well-defined carrier frequency, we get $\omega\tau_{d} \gg 1$, inline with Eq. (\ref{eq:pulse-duration-condition}). In this regime, $\hat{f}_{0}(2\omega\tau_{d})$ in Eq. (\ref{eq:f-hat-omega-tau_d}) almost vanishes, so that the functional $s_{1}$ can be approximated as
\begin{equation}\label{eq:s1_asympt_and_rwa_and_notodd}
	s_{1} \simeq \frac{1}{2}e^{i\phi} \hat{f}_{0}(0).
\end{equation}
%Note that this does not hold when $\hat{f}_{0}(0) = 0$, which is the case, for instance, for a pulse shape with an odd parity. In this case, $\hat{f}_{0}(2\omega\tau_{d})$ may still need to be included.
%
%The term $\tilde{A}_{I}^{(1,1)}(t)$ can be written as
%$\sqrt{n}g\tau_{d} \ll 1$.
%is order of $k-1$ in $a$ and $a^{\dagger}$.
%with
%\begin{equation*}
%	\tilde{A}_{I}^{(1,k)}(t) = \int_{-T_{u}}^{u}{du' f(u')\, i^{k}\,\mathrm{ad}_{\tilde{H}_{g}}^{k} x_{\omega}(u')}, 
%%			\tilde{H}_{g}/i], ..., \tilde{H}_{g}/i]},
%%	\tilde{A}_{I}^{(1,k)}(t) = \int_{-T_{u}}^{u}{du' f(u') [...[x_{\omega}(u'), \tilde{H}_{g}/i], ..., \tilde{H}_{g}/i]},
%\end{equation*}
%for $\tilde{H}_{g} \equiv H_{g} / \hbar g$.
%$\mathrm{ad}_{X}Y \equiv [X,Y]$, $\mathrm{ad}_{X}^{k+1} = \mathrm{ad}_{X}^{k}\mathrm{ad}_{X}$.

From Eqs. (\ref{eq:mathcal-U-magnus}), (\ref{eq:A_I_m_t}) and (\ref{eq:A_I_1_t_decomposed}), we get the leading-order term of $\mathcal{U}$,
\begin{subequations}\label{eq:U_1_t}
\begin{align}
	\mathcal{U}(t) 
	& = \exp[-i A_{I}(t)]\\
%	& = \exp[-i \sum_{m=1}^{\infty}A_{I}^{(m)}(t)]\\
%	& = \exp[-iA_{I}^{(1)}(t) + \mathcal{O}[(\sqrt{n}\Omega\tau_{d})^{2}]]\\
	& \simeq \exp[-iA_{I}^{(1)}(t)]\\
%	& = \exp[-iA_{I}^{(1,0)}(t) + \mathcal{O}[(\sqrt{n}\Omega\tau_{d})(\sqrt{n}g\tau_{d})]]\\
	& \simeq \exp[-iA_{I}^{(1,0)}(t)]\\
	& \equiv U_{1}(t),
\end{align}
\end{subequations}
where the second line holds for $\sqrt{n}\Omega\tau_{d} \ll 1$ and the third line for $\sqrt{n}g\tau_{d} \ll 1$. The $U_{1}(t)$ denotes the leading-order term. From Eq. (\ref{eq:tilde_A_I_1_0_t_explicit}), one can show that the leading term $U_{1}$ is a displacement operator,
\begin{equation*}
	U_{1}(t) = D[z(t)],
\end{equation*}
where the complex amplitude of the displacement can be written as
\begin{equation}\label{eq:zt}
	z(t) = -i\Omega\tau_{d} s_{1}^{*}(t).
\end{equation}
The phase, or direction, of the displacement can be controlled by the carrier-envelope offset phase $\phi$, as can be seen from 
%Eq. (\ref{eq:ft}). This comes from 
Eq. (\ref{eq:s1_asympt_and_rwa_and_notodd}).
%, or more generally, from Eq. (\ref{eq:f-hat-omega-tau_d}).

\begin{figure}
	\centering
	\includegraphics[width=1\linewidth]{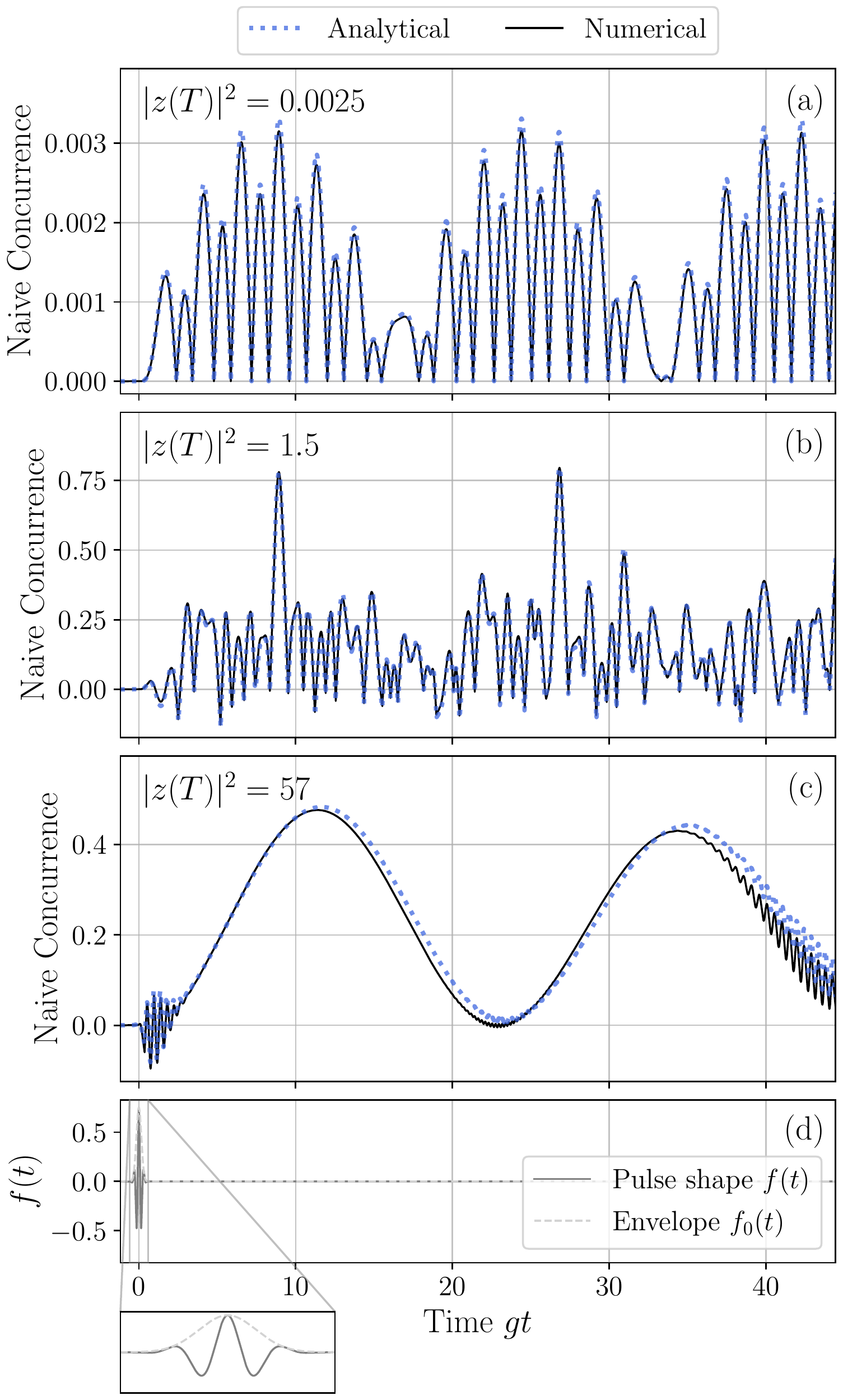}
	\caption{
		Naive concurrence induced by a subcycle pulse with different driving strengths $\Omega$, given by 
%		(a) $\Omega = 0.017\omega$, (b) $\Omega = 0.41\omega$ and (c) $\Omega = 2.5\omega$.
		(a) $\Omega\tau_{d} = 0.0531$, (b) $\Omega\tau_{d} = 1.29$ and (c) $\Omega\tau_{d} = 8$. The pulse duration, envelope, carrier-envelope phase and the cavity-qubit coupling are fixed as $\tau_{d} = \pi / \omega$, $f_{0}(t) = f_{\mathrm{HG},0}(t)$, $\phi = 0$ and $g = 0.05\omega$, respectively. The black solid lines (blue dotted lines) represent numerically (analytically) evaluated naive concurrences. The analytical results are obtained essentially by approximating the time-evolution operator by a displacement operator, i.e. $\mathcal{U}(t)\simeq D[z(t)]$ based on Eq. (\ref{eq:U_1_t}). The displacement amplitude $z(t)$ is given by Eq. (\ref{eq:zt}). In (a)-(c), the corresponding average photon numbers $|z(t)|^{2}$ at the end of the pulse $t=T$ are indicated in each panel. The gray solid line and the light gray dashed line in (d) show the pulse shape $f(t)$ and its envelope $f_{0}(t)$, respectively.
	}
	\label{fig:naive-concurr-U1-dominant}
\end{figure}

%\begin{figure}
%	\centering
%	\includegraphics[width=1\linewidth]{fig/naive-concurr-CHECK_CONVERGENCE_OF_NUMERICAL_SOLUTION-test-106}
%	\caption{
%		(a) Naive concurrence induced by a subcycle pulse. The cavity-qubit coupling $g$, the normalized pulse duration $g\tau_{d}$ and the pulse area $\Omega\tau_{d}$ are indicated on top of panel (a). The black solid line represents a numerical solution. The blue dashed line corresponds to an analytical solution where the effect of the pulse through the time-evolution operator $\mathcal{U}$ is approximated as a displacement $D[z]$ of the cavity mode and a rotation $R[\theta,\mathbf{n}]$ of the qubits, with the amplitude $z$ given as Eq. (\ref{eq:zt}) whereas the rotational angle $\theta$ and the axis $\mathbf{n}$ given as Eq. (\ref{eq:theta-and-nvec}). The values of $|z|^{2}$ and $\theta$ provided in the legend are asymptotic values after the pulse, i.e. $t \gg \tau_{d}$. See text for further details.
%		(b) The pulse shape $f(t)$ is given as Eq. (\ref{eq:ft}). The zero-th order Hermite-Gaussian function $f_{\mathrm{HG},0}(t)$ is used as the envelope function $f_{0}(t)$. The carrier-envelope offset phase is set to $\phi = 0$.
%		%		with its envelope function $f_{0}(t)$ given as Eq. (\ref{eq:ft}).
%	}
%	\label{fig:naive-concurr-checkconvergenceofnumericalsolution-U1-dominant}
%\end{figure}

If $z(t) \simeq 0$ after the pulse, i.e. $t \gg \tau_{d}$, the leading-order term $U_{1}$ has only a transient modulation during the pulse. In order to drive the cavity mode into a coherent state with a nonzero amplitude $z$ after the pulse, a pulse with $s_{1}(t) \neq 0$ for $t \gg \tau_{d}$ is required. With the asymptotic expression for $s_{1}$ given by Eq. (\ref{eq:s1_asympt_and_rwa_and_notodd}), this requires an envelope $f_{0}(t)$ such that $\hat{f}_{0}(0) = \int_{-\infty}^{\infty} dt\,f_{0}(t) \neq 0$. For example, a Gaussian shape $f_{\mathrm{HG},0}(t)$ can be used. % $f_{0}(t) = f_{$\mathrm{HG}$,0}(t)$.
%The leading-order term $U_{1}(t)$ is present 
The concurrence induced by a subcycle pulse with such a shape is shown in Fig. \ref{fig:naive-concurr-U1-dominant}.
%\ref{fig:naive-concurr-checkconvergenceofnumericalsolution-U1-dominant}. 

%\begin{figure}
%	\centering
%	\includegraphics[width=1\linewidth]{fig/naive-concurr-CHECK_CONVERGENCE_OF_NUMERICAL_SOLUTION-test-110-dt_num-0.000628-rotang-6.96e-09-nx--1-ny--5.9e-06-nz-0}
%	\caption{
%		Same as Fig. \ref{fig:naive-concurr-checkconvergenceofnumericalsolution-U1-dominant}, except that the driving strength $\Omega$ is set such that the average number of photons which are pumped into the cavity mode is around one, i.e. $|z|^{2} \sim 1$. In this calculation, $|z|^{2} = 1.7$ photons in average are pumped into the cavity mode.
%	}
%	\label{fig:naive-concurr-U1-dominant-n_avg-near-one}
%\end{figure}

%\begin{figure}
%	\centering
%	\includegraphics[width=1\linewidth]{fig/naive-concurr-CHECK_CONVERGENCE_OF_NUMERICAL_SOLUTION-test-111-dt_num-0.000628-rotang-3.98e-08-nx--1-ny--5.9e-06-nz-0}
%	\caption{
%		Same as Fig. \ref{fig:naive-concurr-checkconvergenceofnumericalsolution-U1-dominant}, except that the driving strength $\Omega$ is set such that the average number of photons which are pumped into the cavity mode is much larger than one, i.e. $|z|^{2} \gg 1$. In this calculation, $|z|^{2} = 57$ photons in average are pumped into the cavity mode.
%	}
%	\label{fig:naive-concurr-U1-dominant-n_avg-large}
%\end{figure}

Depending on the pulse area $\Omega\tau_{d}$, characteristics of the time-dependent concurrence vary. 
For example, for a small pulse area, i.e. $\Omega \tau_{d} \ll 1$, the average number of quanta pumped into the cavity, which is given by $|z(T)|^{2}$, is much smaller than one. Thus, the concurrence is dominated by the interference between states belonging to few-quanta subspaces. This can be seen in Fig. \ref{fig:naive-concurr-U1-dominant}(a).
%in an interference between states belonging to the zero-quanta subspace with those to the single- and two-quanta space.
%For example, concurrence about $0.8$ can be achieved if $\Omega \tau_{d} \sim 1$ such that the generated coherent state has an average photon number of around one, i.e. $|z|^{2} \sim 1$, which is shown in Fig. \ref{fig:naive-concurr-U1-dominant}(b). 
When $\Omega \tau_{d} \sim 1$ such that the generated coherent state has an average photon number of around one, i.e. $|z|^{2} \sim 1$, a value of concurrence larger than $0.75$ can be achieved, which is shown in Fig. \ref{fig:naive-concurr-U1-dominant}(b).
This is consistent with the case of a Fock state where the single-photon state allows to achieve maximal entanglement. 
When many photons are pumped into the cavity mode, so that $|z|^{2} \gg 1$, a smooth oscillation of concurrence appears, as shown in Fig. \ref{fig:naive-concurr-U1-dominant}(c). This is consistent with the concurrence generated by a strong coherent state \cite{Jarvis_2009}.

\subsection{\label{sec:second-leading-order-effect-rotation}The second-order effect: rotation of the qubits}

%Since the leading-order term $U_{1}(t)$ is not exactly the same as $\mathcal{U}(t)$, one needs a correction if a more accurate understanding is required. In addition, 
The propagator $U_{1}(t)$ is a good approximation of $\mathcal{U}(t)$ only when $\sqrt{n}\Omega\tau_{d} \ll 1$ and $\sqrt{n}g\tau_{d} \ll 1$. Thus, even if the pulse is short, satisfying the latter condition, if the pulse area $\sqrt{n}\Omega\tau_{d}$ is not small enough, $U_{1}(t)$ may not be sufficient to describe the dynamics. This is because the series Eq. (\ref{eq:magnus-exponent}) may diverge for a large $\sqrt{n}\Omega\tau_{d}$, in which even the inclusion of high-order terms may not work.
%even if many terms are included.

%The second-leading order is needed for 
In order to describe the case where $\sqrt{n}\Omega\tau_{d}$ is not too small, we proceed with the following decomposition of $\mathcal{U}$:
\begin{equation}\label{eq:mathcal-U_2-def}
	\mathcal{U}(t) \equiv U_{1}(t)\,\mathcal{U}_{2}(t).
\end{equation}
Note that Eq. (\ref{eq:mathcal-U_2-def}) holds for any finite value of $\sqrt{n}\Omega\tau_{d}$, as it is merely a definition of another interaction picture in which the contribution of the leading-order effect $U_{1}(t)$ is subtracted.
%a time-evolution operation in a new interaction picture.
In order to evaluate $\mathcal{U}_{2}(t)$, we find the Hamiltonian which generates $\mathcal{U}_{2}(t)$. Let us denote the Hamiltonian as $H_{II}(t)$. Then the time-evolution operator $\mathcal{U}_{2}(t)$ satisfies $\dot{\mathcal{U}}_{2}(t) = (-i/\hbar)H_{II}(t)\,\mathcal{U}_{2}(t)$. From Eq. (\ref{eq:mathcal-U_2-def}), we get the Hamiltonian,
\begin{equation}\label{eq:H_II_t}
	H_{II}(t) = U_{1}^{\dagger}(t) [ H_{I}(t) - H_{1}(t)] U_{1}(t),
\end{equation}
where $H_{I}(t)$ and $H_{1}(t)$ are the Hamiltonians generating $\mathcal{U}(t)$ and $U_{1}(t)$, respectively. $H_{I}(t)$ can be evaluated from Eqs. (\ref{eq:H_I_t}) and (\ref{eq:tilde_H_I_t}). $H_{1}(t)$ can be obtained by differentiating $U_{1}(t)$ and using the definition of $H_{1}(t)$, namely $\dot{U}_{1}(t) = (-i/\hbar)H_{1}(t)\,U_{1}(t)$. The derivative of $U_{1}(t)$ can be obtained by using the Zassenhaus formula \cite{Magnus} or 
\begin{equation*}
	\frac{d}{dt}{e^{A(t)}} = \int_{0}^{1}{ds \, e^{sA(t)} \, \frac{dA}{dt} \, e^{-sA(t)}} e^{A(t)},
\end{equation*}
which is shown in, e.g., Ref. \cite{PhysRev.84.108}. We can then get the Hamiltonian $H_{1}(t)$ as
\begin{equation}\label{eq:H_1_t}
	H_{1}(t) = H_{e}(t) - \frac{1}{2}\langle z(t) | H_{e}(t) | z(t) \rangle,
\end{equation}
where $|z(t)\rangle \equiv D[z(t)]|0\rangle$ is a coherent state with the amplitude $z(t)$ given as Eq. (\ref{eq:zt}). Inserting Eqs. (\ref{eq:H_I_t}) and (\ref{eq:H_1_t}) into the expression for $H_{II}(t)$ in Eq. (\ref{eq:H_II_t}), we get
\begin{equation*}
	H_{II}(t) = H_{II}^{\prime}(t) + H_{II,z}(t),
\end{equation*}
where
\begin{subequations}
\begin{align}
	\label{eq:H_II_prime_t}
	H_{II}^{\prime}(t) & \equiv U_{1}^{\dagger}(t) [ U_{g}^{\dagger}(t) H_{e}(t) U_{g}(t) - H_{e}(t)] U_{1}(t),\\
	H_{II,z}(t) & \equiv \frac{1}{2}\langle z(t) | H_{e}(t) | z(t) \rangle.
\end{align}
\end{subequations}
Since $H_{II,z}(t)$ is a scalar, it can be subtracted from $H_{II}(t)$ by the transformation
\begin{equation}\label{eq:mathcal_U_2_with_U_II_z_and_U_2_prime}
	\mathcal{U}_{2} \equiv U_{II,z}(t) \, \mathcal{U}_{2}^{\prime}(t),
\end{equation}
with
\begin{equation*}
	U_{II,z}(t) = \exp\left[-\frac{i}{\hbar} \int_{-T}^{t}{dt' H_{II,z}(t')}\right].
\end{equation*}
We get then the differential equation
\begin{equation*}
	\dot{\mathcal{U}}_{2}^{\prime}(t) = - \frac{i}{\hbar} H_{II}^{\prime}(t) \, \mathcal{U}_{2}^{\prime}(t),
\end{equation*}
whose formal solution is again given by the Magnus expansion,
\begin{subequations}\label{eq:mathcal_U_2_prime_magnus}
\begin{align}
	\label{eq:U_2_prime_t}
	\mathcal{U}_{2}^{\prime}(t) & \equiv \exp[-i A'_{II}(t)],\\
	\label{eq:A_II_t_magnus}
	A_{II}^{\prime}(t) & = \sum_{m=1}^{\infty}{ A_{II}^{\prime (m)}(t) }.
\end{align}
\end{subequations}
However, the order of magnitude of each term is different from that of $A_{I}^{(m)}(t)$ in Eq. (\ref{eq:magnus-exponent}). In Appendix \ref{sec:magnus-expansion-order-of-magnitudes}, we show that 
\begin{equation}\label{eq:A_II_prime_m_t_in_tilde_Omega_n_tilde_g_n}
	A_{II}^{\prime (m)}(t) = \mathcal{O}[(\tilde{\Omega}_{n}\tilde{g}_{n})^{m}].
\end{equation}
It has an additional factor, $\tilde{g}_{n} \equiv \sqrt{n}g\tau_{d}$, compared to the previous case, $A_{I}^{(m)}(t) = \mathcal{O}[(\tilde{\Omega}_{n})^{m}]$ in Eq. (\ref{eq:A_I_m_t_in_tilde_Omega_n}). This arises from a property of ${H}_{II}^{\prime}(t)$, given by Eq. (\ref{eq:H_II_prime_t}), where $H_{e}(t)$ is subtracted from $U_{g}^{\dagger}(t)H_{e}(t)U_{g}(t)$. Applying the identity (\ref{eq:campbell}) to $U_{g}^{\dagger}(t)H_{e}(t)U_{g}(t)$, the Hamiltonian can be expanded as
\begin{equation}\label{eq:tilde_H_II_t_expanded_in_g_tau_d}
\begin{split}
	\tilde{H}_{II}^{\prime}(t)
	\equiv&\,{H}_{II}^{\prime}(t) / \hbar\Omega\\
	=&\,(g\tau_{d})^{1} f(u) u U_{1}^{\dagger}(u) [i\tilde{H}_{g},x_{\omega}(u)] U_{1}(u)\\
	+&\,\mathcal{O}[(g\tau_{d})^{2}].
\end{split}
\end{equation}
We note that $\tilde{H}_{II}^{\prime}(t) = \mathcal{O}[(g\tau_{d})^{1}]$, whereas $\tilde{H}_{I}(t) = \mathcal{O}[(g\tau_{d})^{0}]$, as can be seen from Eq. (\ref{eq:tilde_H_I_g-tau_d-expansion}).
%If we compare this with Eq. (\ref{eq:tilde_H_I_g-tau_d-expansion}), we can see that 

For $\tilde{\Omega}_{n}\tilde{g}_{n} \ll 1$, the dominant term in the expansion Eq. (\ref{eq:A_II_t_magnus}) is $A_{II}^{\prime(1)}(t)$, cf. Eq. (\ref{eq:A_II_prime_m_t_in_tilde_Omega_n_tilde_g_n}). This condition allows us to use a pulse such that $1 < \tilde{\Omega}_{n} \ll (\tilde{g}_{n})^{-1}$ by using a sufficiently short pulse $\tilde{g}_{n} \ll 1$. Note that if we use the Magnus expansion in Eq. (\ref{eq:magnus-exponent}) to get the second- and higher-order terms, the pulse area has to be limited by $\tilde{\Omega}_{n} \ll 1$ to assure the expansion convergence. By shifting to the second interaction picture, Eq. (\ref{eq:U-decomposed}), we can use a pulse with a larger $\tilde{\Omega}_{n}$. This is required to generate a coherent state with amplitude much larger than $1$ since $z(t) = \mathcal{O}[\Omega\tau_{d}]$. Phenomena such as the collapse and revival of qubits observables are visible only in this regime \cite{PhysRevLett.44.1323}.
%if the pulse duration is sufficiently short to satisfies 

Using Eq. (\ref{eq:tilde_H_II_t_expanded_in_g_tau_d}) to evaluate $A_{II}^{\prime(1)}(t)$, we get
\begin{equation}\label{eq:A_II_prime_1_t_decomposed}
	A_{II}^{\prime(1)}(t) = A_{II}^{\prime(1,1)}(t) + \mathcal{O}[\tilde{\Omega}_{n}\tilde{g}_{n}^{2}].
\end{equation}
The leading term is given as
\begin{equation}\label{eq:tilde_A_II_prime_1_1_t}
\begin{split}
	\tilde{A}_{II}^{\prime(1,1)}(t) 
	& \equiv {A}_{II}^{\prime(1,1)}(t) / (\Omega\tau_{d})(g\tau_{d})\\
	& = s_{(1,1)}(t) (-i\sigma^{-}) + s_{(1,1)}^{*}(t) (i\sigma^{+}),
\end{split}
\end{equation}
where 
\begin{equation*}
	s_{(1,1)}(t) = \int_{-T/\tau_{d}}^{t/\tau_{d}}{du \, u f(u) e^{-i \omega \tau_{d} u} }.
\end{equation*}
Similar to $s_{1}$ in Eq. (\ref{eq:s1_asympt}), when $t \gg \tau_{d}$, $s_{(1,1)}(t)$ can be approximated as
\begin{equation*}
	s_{(1,1)} \simeq i \hat{f}^{(1)}(\omega\tau_{d}),
\end{equation*}
where $\hat{f}^{(1)} \equiv d\hat{f}/dk$ is the derivative of the Fourier transform $\hat{f}$ of $f$. 
In our case, the pulse has a well-defined carrier frequency $\omega$, entering Eq. (\ref{eq:ft}), which means $\omega \tau_{d} \gg 1$. Therefore, the functional can be approximated as
\begin{equation}\label{eq:s_1_1_large-omega-tau_d}
	s_{(1,1)} \simeq \frac{1}{2} e^{i\phi} i \hat{f}^{(1)}_{0}(0),
\end{equation}
where $\hat{f}^{(1)}_{0} \equiv d\hat{f}_{0}/dk$. Note that for any envelope $f_{0}(u)$ with even parity, $\hat{f}^{(1)}_{0}(0) = 0$.

From Eqs. (\ref{eq:mathcal_U_2_prime_magnus}), (\ref{eq:A_II_prime_m_t_in_tilde_Omega_n_tilde_g_n}) and (\ref{eq:A_II_prime_1_t_decomposed}), we now get the second leading-order term as
%is given as
\begin{subequations}
\begin{align}
	\mathcal{U}^{\prime}_{2}(t) 
	& = \exp[-iA_{II}^{\prime}(t)]\\
	& \simeq \exp[-iA_{II}^{\prime(1)}(t)]\\
	& \simeq \exp[-iA_{II}^{\prime(1,1)}(t)]\\
	& \equiv U_{2}(t),
\end{align}
\end{subequations}
where the second line holds for $\tilde{\Omega}_{n}\tilde{g}_{n} \ll 1$ and the third line for $\tilde{g}_{n} \ll 1$. The last line defines the second-leading term $U_{2}(t)$. With Eq. (\ref{eq:tilde_A_II_prime_1_1_t}), one can show that $U_{2}(t)$ is a rotational operator:
\begin{equation*}
	U_{2}(t) = R[\theta(t);\mathbf{n}(t)],
\end{equation*}
where $R[\theta;\mathbf{n}] \equiv \exp[-i \theta \mathbf{n}\cdot \pmb{\sigma} / 2]$ for an angle $\theta$, a rotational axis $\mathbf{n}$ of unit length and $\pmb{\sigma} = (\sigma^{x}, \sigma^{y}, \sigma^{z})$ with $\sigma^{j} \equiv \sigma_{A}^{j} + \sigma_{B}^{j}$ for $j \in \{x,y,z\}$.
It rotates the Bloch vector of each qubit. 
%From Eq. (\ref{eq:A_II_prime_1_k_t}) and (\ref{eq:tilde_A_II_prime_1_1_t}), 
The angle and the rotational axis are given as
%\begin{subequations}
\begin{subequations}\label{eq:theta-and-nvec}
\begin{align}
	\label{eq:theta-U2}
	\theta(t) & = 2 (\Omega\tau_{d})(g\tau_{d}) |s_{(1,1)}(t) |,\\
	\label{eq:nvec-U2}
	\mathbf{n}(t) & = \big(\sin[\phi_{(1,1)}(t)], -\cos[\phi_{(1,1)}(t)], 0\big),
\end{align}
\end{subequations}
%\end{subequations}
for $s_{(1,1)}(t) \equiv |s_{(1,1)}(t)| e^{i\phi_{(1,1)}(t)}$.

%\subsection{Concurrence depending on the pulse shape}

\begin{figure}
	\centering
	\includegraphics[width=1\linewidth]{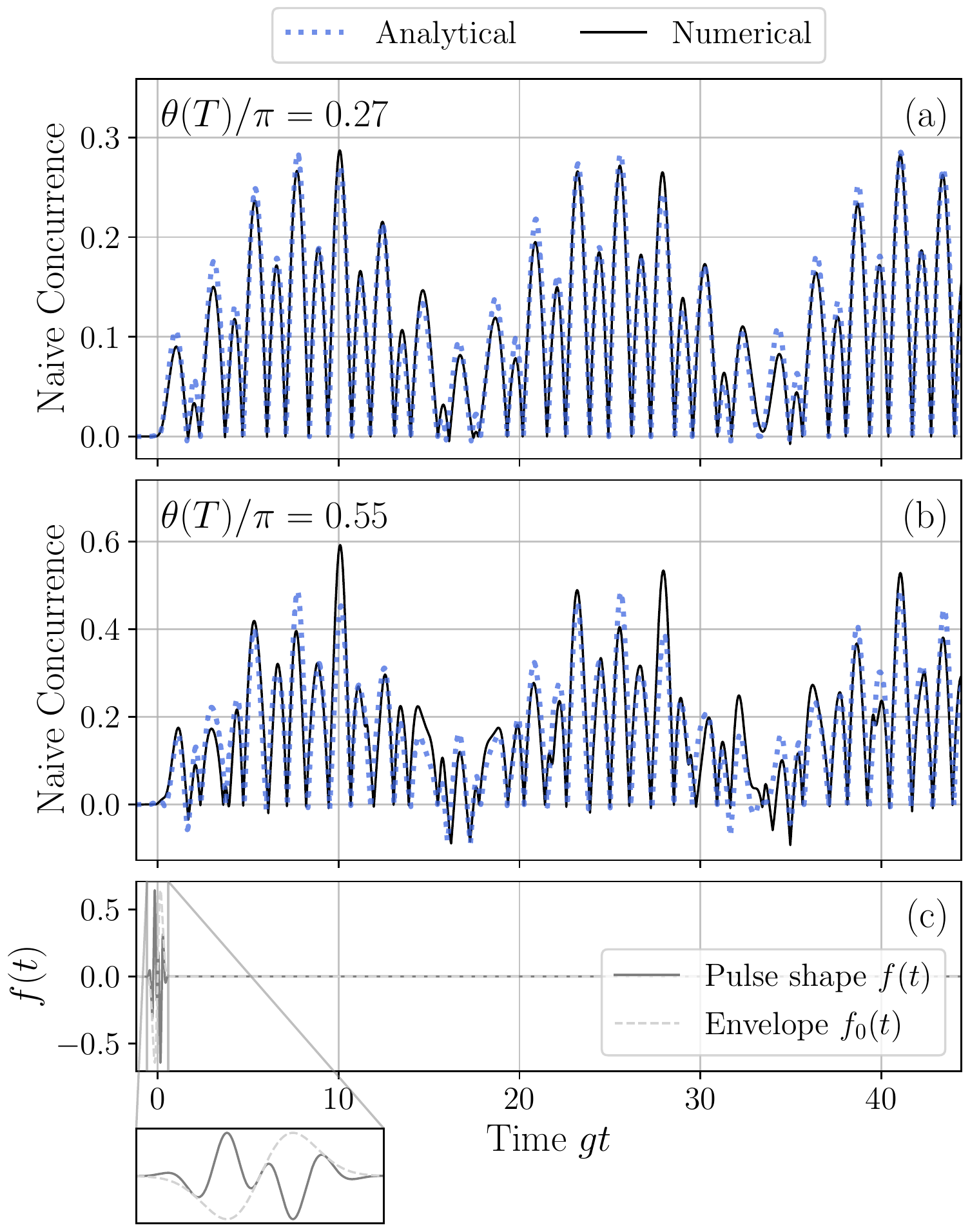}
	\caption{
		Same as Fig. \ref{fig:naive-concurr-U1-dominant}, except that the pulse envelope is given by $f_{0}(t) = f_{\mathrm{HG},1}(t)$ and the driving strength $\Omega$ is given by (a) $\Omega \tau_{d} = 2.05$ and (b) $\Omega \tau_{d} = 4.1$. The analytical solution is based on the approximation $\mathcal{U}(t) \simeq D[z(t)]R[\theta(t);\mathbf{n}(t)]$, i.e. including the terms up to the second order. The expressions for the displacement amplitude $z(t)$, the rotational angle $\theta(t)$ and the rotational axis $\mathbf{n}(t)$ are given by Eqs. (\ref{eq:zt}), (\ref{eq:theta-U2}) and (\ref{eq:nvec-U2}), respectively. For the given envelope, the displacement amplitude almost vanishes after the pulse, i.e. at $t=T$. The rotational angle at the end of the pulse, $\theta(T)$, is indicated in panels (a) and (b).
%		is given by Eq. (\ref{eq:mathcal-U_2-def}).
		}
	\label{fig:naive-concurr-rotation-dominant}
\end{figure}

%\begin{figure}
%	\centering
%	\includegraphics[width=1\linewidth]{fig/naive-concurr-CHECK_CONVERGENCE_OF_NUMERICAL_SOLUTION-test-105}
%	\caption{Same as Fig. \ref{fig:naive-concurr-checkconvergenceofnumericalsolution-U1-dominant}, except that the first order Hermite-Gaussian function $f_{\mathrm{HG},1}(t)$ is used as the envelope $f_{0}(t)$.}
%	\label{fig:naive-concurr-checkconvergenceofnumericalsolution-U2-dominant}
%\end{figure}

In order to have a nonzero rotation after the pulse, i.e. $\theta(t) \neq 0$ for $t \gg \tau_{d}$, we need a pulse envelope $f_{0}(t)$ with $s_{(1,1)}(t) \neq 0$ for $t \gg \tau_{d}$, as follows from Eq. (\ref{eq:theta-and-nvec}). Using Eq. (\ref{eq:s_1_1_large-omega-tau_d}), the condition reads $i\hat{f}_{0}(0) = \int_{-\infty}^{\infty} dt \, t f_{0}(t) \neq 0$. Thus, if the pulse envelope is even, the rotational angle essentially vanishes. 
If we use an odd envelope, a nonzero rotation is possible. In this case, the leading-order contribution $U_{1}$ is turned off after the pulse since $s_{1}(t) \simeq 0$ for $t \gg \tau_{d}$. 
The concurrence induced by a subcycle pulse with an odd envelope is shown in Fig. \ref{fig:naive-concurr-rotation-dominant}.
%The concurrence induced by a subcycle pulse is shown in Fig. \ref{fig:naive-concurr-checkconvergenceofnumericalsolution-U1-dominant} and \ref{fig:naive-concurr-checkconvergenceofnumericalsolution-U2-dominant}. 
%
Comparing Fig. \ref{fig:naive-concurr-rotation-dominant} with Fig. \ref{fig:naive-concurr-U1-dominant}, the cavity-qubit coupling $g$, the normalized pulse duration $g\tau_{d}$ and the pulse area $\Omega\tau_{d}$ are the same in both figures. The only difference is the shape of the pulse envelope $f_{0}(t)$, resulting in qualitatively distinct dynamics of the concurrence. 
%
%In Fig. \ref{fig:naive-concurr-checkconvergenceofnumericalsolution-U1-dominant} and \ref{fig:naive-concurr-checkconvergenceofnumericalsolution-U2-dominant}, $f_{\mathrm{HG},0}(t)$ and  $f_{\mathrm{HG},1}(t)$ are used as the envelope function, respectively. Both Hermite-Gaussian functions are given as Eq. (\ref{eq:hg_m}). 
%Both results seem to be qualitatively different, depending on the pulse shape. In this section, we show that the effect of the pulse can be approximated as a displacement $D[z]$ of the cavity mode and a rotation $R[\theta;\mathbf{n}]$ of the qubits with a proper selection of the displacing amplitude $z$, the rotating angle $\theta$ and the rotational axis $\mathbf{n}$. Explicit expressions of all quantities will be given as a functional of the pulse shape $f(t)$ as well as other parameters such as $g$, $\tau_{d}$ and $\Omega$.

%\begin{figure}
%	\centering
%	\includegraphics[width=1\linewidth]{fig/naive-concurr-CHECK_CONVERGENCE_OF_NUMERICAL_SOLUTION-test-113-dt_num-0.000628-rotang-1.71-nx-9.2e-13-ny--1-nz-0}
%	\caption{Same as Fig. \ref{fig:naive-concurr-checkconvergenceofnumericalsolution-U2-dominant}, except that the driving strength $\Omega$ is stronger, such that $\theta = 0.55\,\pi$.}
%	\label{fig:naive-concurr-U2-dominant-stronger}
%\end{figure}

A stronger driving results in an increased angle of rotation. An angle larger than $\pi/2$ is achieved in Fig. \ref{fig:naive-concurr-rotation-dominant}(b). 
%In Fig. \ref{fig:naive-concurr-U2-dominant-stronger}(a), 
There, the numerical result shows that there is an additional effect of the pulse on top of the displacement $U_{1} = D[z]$ and the rotation $U_{2} = R[\theta,\mathbf{n}]$. Those additional corrections can be attributed to higher-order terms which are treated in the following section.

\subsection{\label{sec:higher-order-effects}Higher-order effects: conditional displacement and rotation around the $z$-axis}

The first two leading-order effects are identified as a displacement of the cavity mode and a rotation of the qubits. One may describe the effect of the pulse approximately with these two operations. 
%This can be a good approximation if the pulse duration is short enough $\sqrt{n}g{\tau}_{d} \ll 1$ and the pulse strength satisfies $$
%However, this is not an exact description. 
%In order to understand how much error 
However, there are higher-order terms which do not vanish in general.
Understanding the higher-order terms might help to find other possible types of operations apart from the displacement or the rotation.
%The higher-order terms can be obtained

As an example, we present two quadratic terms in $g\tau_{d}$. They can be found by expanding the exponent $A_{II}^{\prime}(t)$, Eq. (\ref{eq:A_II_t_magnus}), in the orders of $\Omega\tau_{d}$ and $g\tau_{d}$. There are only two such terms, $A_{II}^{\prime(1,2)}(t)$ and $A_{II}^{\prime(2,2)}(t)$, which are quadratic in $g\tau_{d}$. The first term ${A}_{II}^{\prime (1,2)}(t)$ generates a conditional displacement where the direction to which the cavity mode is displaced depends on the state of the qubits. To be explicit,
\begin{equation*}
\begin{split}
	\tilde{A}_{II}^{\prime (1,2)}(t)
	& \equiv {A}_{II}^{\prime (1,2)}(t) / (\Omega\tau_{d})(g\tau_{d})^{2}\\
	& = \sigma^{z} [ s_{(1,2)}(t) a + s_{(1,2)}^{*}(t) a^{\dagger} ].
\end{split}
\end{equation*}
The functional $s_{(1,2)}(t)$ modulates the displacement amplitude and is given as
\begin{equation*}
	s_{(1,2)}(t) = \int_{-T/\tau_{d}}^{t/\tau_{d}}{
		du \frac{u^{2}}{2!} f(u) e^{-i \omega \tau_{d} u}
	}.
\end{equation*}
The second term $A_{II}^{\prime(2,2)}(t)$ represents a rotation of qubits around the $z$-axis,
\begin{equation*}
\begin{split}
	\tilde{A}_{II}^{\prime (2,2)}(t)
	& \equiv {A}_{II}^{\prime (2,2)}(t) / (\Omega\tau_{d})^{2}(g\tau_{d})^{2}\\
	& = \sigma^{z} [ s_{(2,2)}(t) + s_{(2,2)}^{*}(t) ].
\end{split}
\end{equation*}
The functional $s_{(2,2)}(t)$ determines the corresponding angle of rotation and can be written as
\begin{equation*}
	s_{(2,2)}(t) = s_{(2,2),1}(t) + s_{(2,2),2}(t),
%	s_{(2,2)}(t) = \int_{-T/\tau_{d}}^{t/\tau_{d}}{
%		du [\frac{u^{2}}{2!} f(u) (-i) s_{1}^{*}(u) + \frac{u}{1!} f(u) (-i) s_{(1,1)}^{*}(u)] e^{-i \omega \tau_{d} u}
%	}.
\end{equation*}
where
\begin{equation*}
\begin{split}
	s_{(2,2),1}(t) & = \int_{-T/\tau_{d}}^{t/\tau_{d}}{
		du \frac{u^{2}}{2!} f(u) (-i) s_{1}^{*}(u) e^{-i \omega \tau_{d} u}
	},\\
	s_{(2,2),2}(t) & = \int_{-T/\tau_{d}}^{t/\tau_{d}}{
		du \frac{u}{1!} f(u) (-i) s_{(1,1)}^{*}(u) e^{-i \omega \tau_{d} u}
	},
\end{split}
\end{equation*}
originate from the first and the second Magnus term generated by $H_{II}^{\prime}(t)$, Eq. (\ref{eq:A_II_t_magnus}), respectively.
%\begin{equation*}
%\begin{split}
%	\tilde{A}_{II}^{\prime (1,2)}(t)
%	& = \sigma^{z}[ s_{(1,2)}(t) a + s_{(1,2)}^{*}(t) a^{\dagger} ]\\
%	& + (\Omega\tau_{d})[s_{(1,2)(1)}(t) + s_{(1,2)(1)}^{*}(t)]\sigma^{z},
%\end{split}
%\end{equation*}
%where
%\begin{subequations}
%\begin{align*}
%	s_{(1,2)}(t) & = \int_{-T/\tau_{d}}^{t/\tau_{d}}{
%		du \frac{u^{2}}{2} f(u) e^{-i \omega \tau_{d} u}
%	}\\
%	s_{(1,2)(1)}(t) & = \int_{-T/\tau_{d}}^{t/\tau_{d}}{
%		du \frac{u^{2}}{2}f(u) (-i) s_{1}^{*}(u) e^{-i \omega \tau_{d} u}
%	}
%\end{align*}
%\end{subequations}
%\begin{subequations}
%\begin{align*}
%	s_{(1,2)}(t) & = \int_{-T/\tau_{d}}^{t/\tau_{d}{du}\\
%	s_{(1,2)(1)}(t) & = 
%\end{align*}
%\end{subequations}

%(derivation of the higher-order, possibly directly from the expression of $A_{I}^{(1,2)}$ and $A_{I}^{(2,2)}$ or other relevant terms.)

\subsection{Towards exact operation}\label{sec:towards-exact-operations}

%\begin{figure}
%	\centering
%	\includegraphics[width=1\linewidth]{fig/fidel-along-pulse-duration-test-043}
%	\caption{}
%	\label{fig:fidel-along-pulse-duration-test-043}
%\end{figure}
%
%
%
%
%\begin{figure}
%	\centering
%	\includegraphics[width=1\linewidth]{fig/fidel-along-pulse-duration-test-039}
%	\caption{}
%	\label{fig:fidel-along-pulse-duration-test-039}
%\end{figure}

\begin{figure}
	\centering
	\includegraphics[width=1\linewidth]{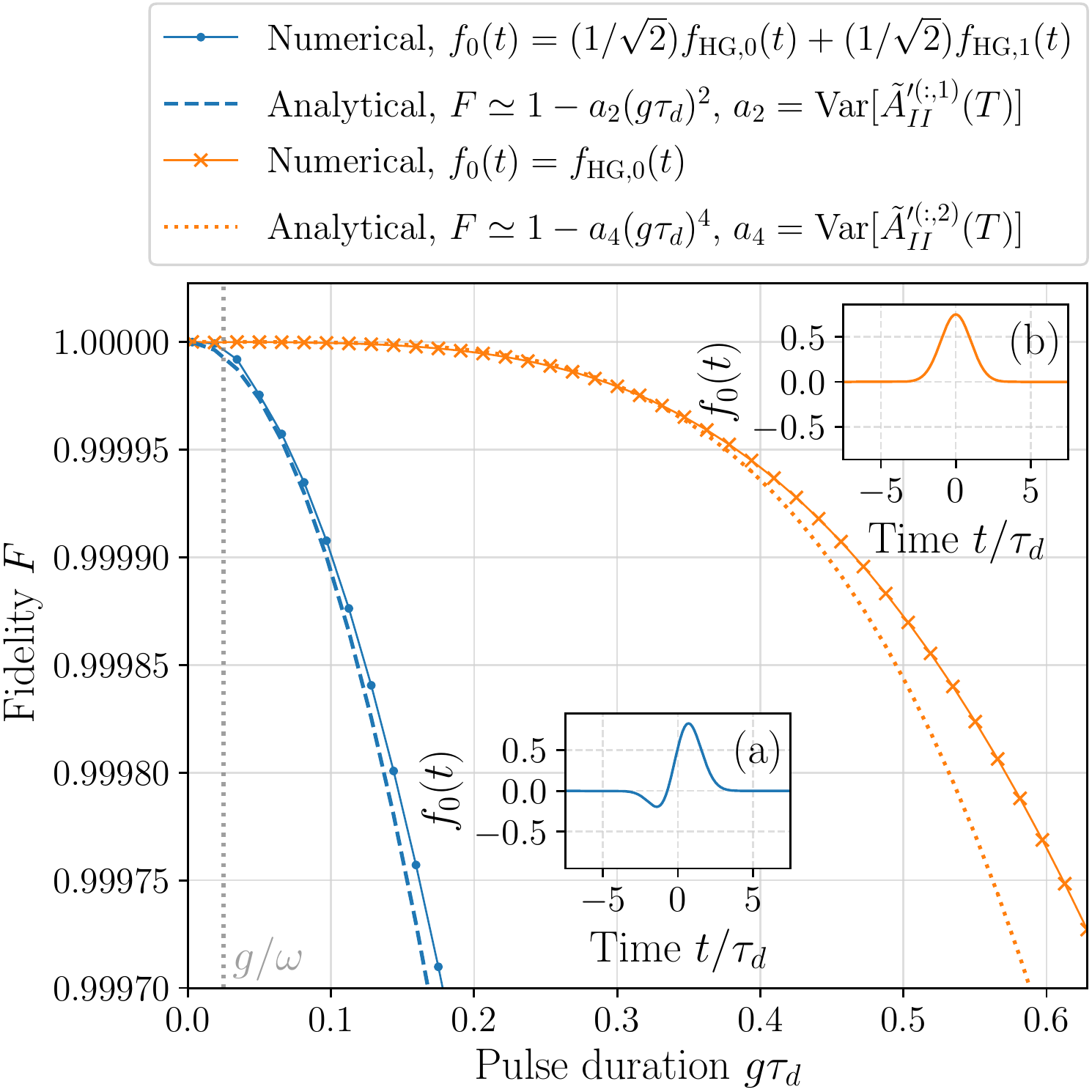}
	\caption{State fidelity for a displacement operation $D[z_{0}]$ with $z_{0} = -0.05i$, implemented by a subcycle pulse. The blue solid line with circles (blue dashed line) indicates numerically (analytically) evaluated fidelity for an envelope shown in inset (a). Likewise, the orange solid line with crosses (orange dotted line) represents numerically (analytically) evaluated fidelity for an envelope shown in inset (b).}
	\label{fig:fidel-comparison-for-different-pulse-shapes}
\end{figure}

Another advantage of the identifying the higher-order terms is improvement of the accuracy of a given operation.
%an ability to implement an operation with a relatively high accuracy.
One can enhance the accuracy of an operation by eliminating irrelevant terms. This can be done by searching a set of the driving parameters, including the pulse shape, which makes those terms vanish. 

As an example, let us discuss how to obtain a displacement with a given amplitude $z_{0}$ with a certain accuracy.
%, one may set the pulse area to 
%In order to implement the given amplitude $z_{0}$, we refer to expression of the displacement amplitude $z(t)$ given as Eq. (\ref{eq:zt}). 
From the displacement amplitude $z(t)$ given by Eq. (\ref{eq:zt})
we require that at the end of the operation, $t = T$, the amplitude of the displacement reaches the desired value $z_{0}$, i.e. $z(T) = z_{0}$. One can indeed find parameters satisfying this requirement. For $T \gg \tau_{d}$ and $\omega \tau_{d} \gg 1$, being valid in the considered regime, we can use Eq. (\ref{eq:s1_asympt_and_rwa_and_notodd}). By selecting a pulse shape such that $\hat{f}(0) > 0$, we obtain:
\begin{subequations}
\begin{align}
	\label{eq:Omega_tau_d_fixed}
	\Omega\tau_{d} & \simeq 2 |z_{0}| / \hat{f}_{0}(0),\\
	\phi & \simeq -\phi_{0} - \pi/2,
\end{align}
\end{subequations}
where $z_{0} \equiv |z_{0}| e^{i\phi_{0}}$.
%From this requirement, we get 
%The error of the operations would depend on the driving parameters.
%Understanding the structure of the error is useful for enhancing the accuracy of the operations
%Satisfying the condition gives a displacement

%The resulting operation is essentially a displacement. 
The error in the resulting operation with respect to the displacement determined by Eq. (\ref{eq:Omega_tau_d_fixed}) can be estimated in terms of the normalized pulse duration $g\tau_{d}$. One can show that as $g\tau_{d} \rightarrow 0$ the error becomes arbitrarily small. In practice, the pulse cannot be infinitesimally short but has a finite duration. For a given finite pulse duration, what is important is the order of the error in terms of $g\tau_{d}$. In order to quantify the error, we use the state fidelity, comparing the target state with the actual state. Starting from the ground state $|\psi(-T)\rangle = |00;0\rangle$ of the system at time $t = -T$, the pulse finishes displacing the cavity mode by $z(T) = z_{0}$ at time $t = T$. In the interaction picture defined by Eq. (\ref{eq:U-decomposed}), the target state, denoted as $|\psi_{0}\rangle = |00;z_{0}\rangle \equiv |00\rangle |z_{0}\rangle$, is a displaced ground state, where $|z_{0}\rangle \equiv D[z_{0}]|0\rangle$ is a coherent state. The state fidelity $F$ is defined as the probability of measuring the target state from the actual state of the system $|\psi(T)\rangle \equiv \mathcal{U}(T) |\psi(-T)\rangle$:
\begin{equation*}
	F = |\langle \psi_{0} | \psi(T) \rangle|^{2}.
\end{equation*}
Using Eq. (\ref{eq:mathcal-U_2-def}) with (\ref{eq:mathcal_U_2_with_U_II_z_and_U_2_prime}) and identifying $U_{1}(T) = D[z(T)] = D[z_{0}]$, the fidelity can be written as
\begin{equation}\label{eq:F-of-mathcal_U_2_prime}
	F = |\langle 00;0 | \,\mathcal{U}_{2}^{\prime}(T) | 00;0 \rangle|^{2}.
\end{equation}
The leading-order term, which is the displacement, cancels out the displacement of the target state and what is left is $\mathcal{U}_{2}^{\prime}(T)$. Thus, the dominant error term for the displacement operation is determined by the leading term of $\mathcal{U}_{2}^{\prime}(T)$, which is $A_{II}^{\prime(1,1)}(T)$.
%, which can be seen from Eq. (\ref{eq:mathcal-U_2-def}). 
For a fixed $\Omega\tau_{d}$ as given in Eq. (\ref{eq:Omega_tau_d_fixed}), the order of magnitude of the error term is
\begin{equation*}
	A_{II}^{\prime(1,1)}(T) = \mathcal{O}[(g\tau_{d})].
\end{equation*}
Note that $A_{II}^{\prime(1,1)}(T)$ is the only term that is linear in $g\tau_{d}$ in the exponent $A_{II}^{\prime}(T)$ of $\mathcal{U}_{2}^{\prime}(T)$.
Let us denote $A_{II}^{\prime(:,1)}(T) = A_{II}^{\prime(1,1)}(T)$, where `$:$' in the superscript implies `all' orders of $\Omega\tau_{d}$ for the given order of $g\tau_{d}$ which is $1$ in this case.
% and since it is linear in $g\tau_{d}$, 
Expanding $\mathcal{U}_{2}^{\prime}(T)$ with respect to $g\tau_{d}$, we arrive at
\begin{equation*}
	F = 1 - (g\tau_{d})^{2} \mathrm{Var}[\tilde{A}_{II}^{\prime(:,1)}] + \mathcal{O}[(g\tau_{d})^{3}],
\end{equation*}
where $\tilde{A}_{II}^{\prime(:,1)}(t) \equiv {A}_{II}^{\prime(:,1)}(t) / (g\tau_{d}) = \tilde{A}_{II}^{\prime(1,1)}(t)$ and the variance is evaluated with respect to the initial state $|\psi(-T)\rangle = |00;0\rangle$.
%From the expression of $\tilde{A}_{II}^{\prime(1,1)}(t)$ in Eq. (\ref{eq:tilde_A_II_prime_1_1_t}), 
%From the previous result on the second-leading order term, 
In general, especially for a pulse envelope without definite parity, $\tilde{A}_{II}^{\prime(1,1)}(T)$ is nonzero, as can be seen from Eq. (\ref{eq:tilde_A_II_prime_1_1_t}). An example of such pulse envelope is $f_{0}(t) = (1/\sqrt{2})f_{\mathrm{HG},0}(t) + (1/\sqrt{2})f_{\mathrm{HG},1}(t)$. However, when the pulse envelope has an even parity, e.g. $f_{0}(t) = f_{\mathrm{HG},0}(t)$, then the leading error term, $\tilde{A}_{II}^{\prime(1,1)}(T)$, almost vanishes, which can be seen from Eqs. (\ref{eq:tilde_A_II_prime_1_1_t}) and (\ref{eq:s_1_1_large-omega-tau_d}). Since the linear term in $g\tau_{d}$ is almost zero, the error is dominated by quadratic terms. As shown in Sec. \ref{sec:higher-order-effects}, there are two quadratic terms in $g\tau_{d}$. Denoting their sum as
\begin{equation*}
\begin{split}
	{A}_{II}^{\prime(:,2)}(t) 
	& = {A}_{II}^{\prime(1,2)}(t) + {A}_{II}^{\prime(2,2)}(t)\\
	& \equiv (g\tau_{d})^{2} \tilde{A}_{II}^{\prime(:,2)}(t),
\end{split}
\end{equation*}
%the terms quadratic in $g\tau_{d}$.
the fidelity can be written as
\begin{equation*}
	F = 1 - (g\tau_{d})^{4} \mathrm{Var}[\tilde{A}_{II}^{\prime(:,2)}] + \mathcal{O}[(g\tau_{d})^{5}].
\end{equation*}
In Fig. \ref{fig:fidel-comparison-for-different-pulse-shapes}, we show the fidelity for $z_{0} = -0.05i$ and for two exemplary shapes of the pulse envelope $f_{0}(t)$. For both pulse shapes, the error converges to zero as the pulse duration becomes shorter. 
For $f_{0}(t) = f_{\mathrm{HG},0}(t)$ the convergence rate is higher. In this case 
%has relatively higher convergence rate. This is because the pulse shape turned off the
the leading term $A_{II}^{\prime(1,1)}(T)$ of the error is suppressed since
%by making 
the functional $s_{(1,1)}(T)$ in Eq. (\ref{eq:tilde_A_II_prime_1_1_t}) almost vanishes.
%The convergence rate of the error is higher for the $f_{\mathrm{HG},0}(t)$ 
A higher convergence rate implies that for a given requirement on the fidelity a longer pulse can be utilized.
%relatively longer, 
This is more desirable for experimental realization, because too short pulses can be problematic both in terms the generation and avoiding certain types of operation errors, as mentioned in the beginning of this section.
%of inducing a desirable action on the system.
%the required shortness of the pulse duration is 
%(condition for reducing the error due to the higher order term.)
One may go beyond the presented convergence rate by eliminating even higher-order error terms $A_{II}^{\prime(:,k)}(T)$ for $k \ge 2$, through the pulse shaping. By this method, one may systematically increase the convergence rate to the extent that a desired operation with a required fidelity can be implemented based on a pulse of available duration.

\section{\label{sec:entanglement-generation-by-quasistatic-driving}Entanglement generation by quasistatic driving, $\tau_{d} \gg T_{g}$}

We consider a quasistatic driving where the duration $\tau_{d}$ of the external driving is longer than the characteristic timescale of the system, $T_{g}$. We first assume that the envelope function is constant, i.e. $f_{0}(t) = 1$. For a fixed driving amplitude $\Omega$, we describe the ground state of the Hamiltonian. Then, we increase the driving strength adiabatically from zero to a finite value, so that the system remains in the ground state corresponding to the instantaneous value of the driving strength at each time moment.

\begin{figure}
	\centering
	\includegraphics[width=1\linewidth]{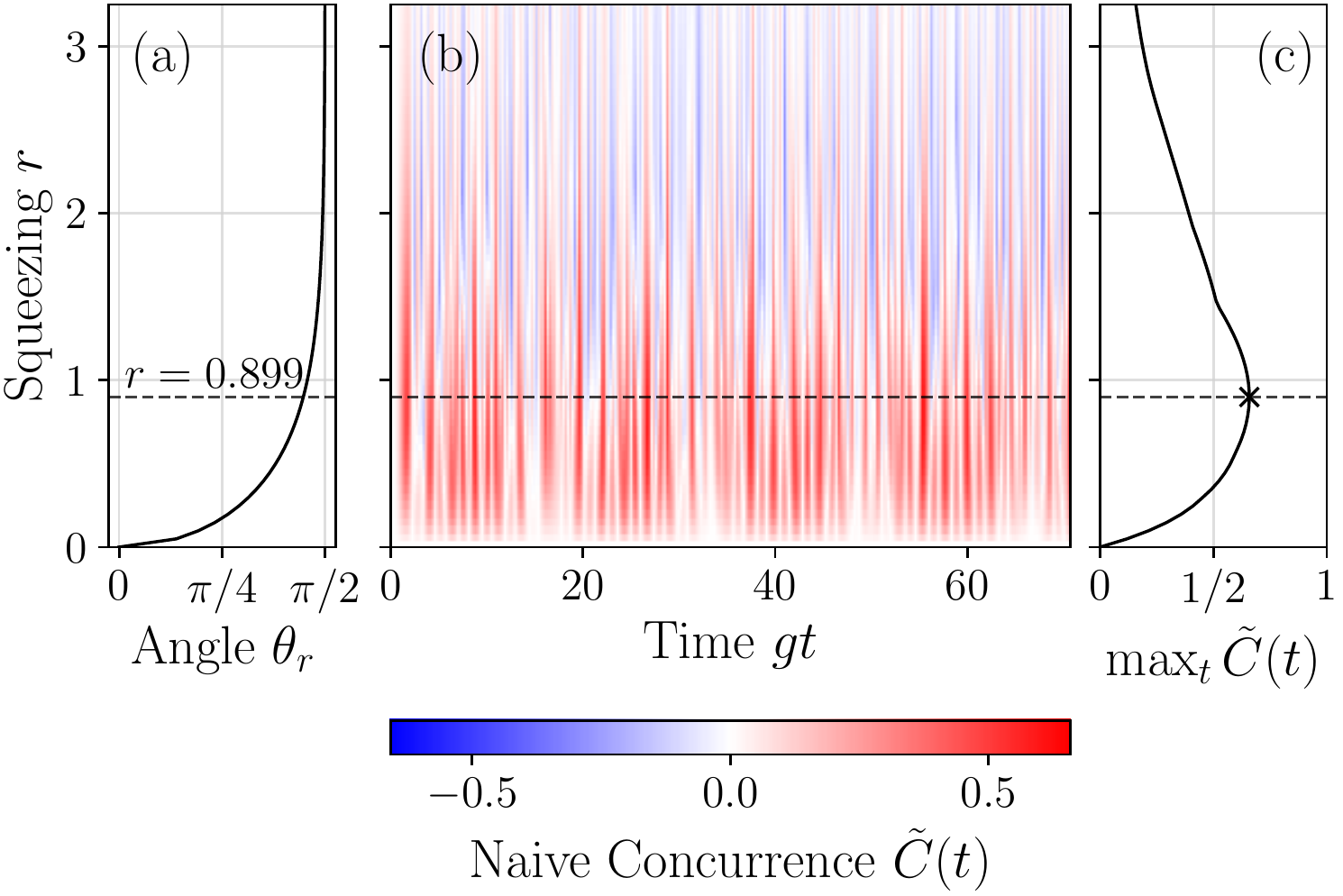}
	\caption{
		(a) The relation between the squeezing parameter $r$ of the cavity mode and the rotational angle $\theta_{r}$ for the qubits. (b) The naive concurrence $\tilde{C}(t)$ versus the squeezing parameter $r$ and time $t$. (c) Naive concurrence maximized with respect to time for each given squeezing $r$.
%		The solid line represents the naive concurrence maximized with respect to time for each given squeezing $r$.
		%The dotted line indicates the average photon number of the squeezed state $|r\rangle$.
	}
	\label{fig:concurrence-along-squeezing}
\end{figure}

For this quasistatic driving, we set $\phi = 0$ and $f_{0}(t) = 1$ in Eq. (\ref{eq:ft}). Notice that a nonzero $\phi$ would correspond to a rotation of both the state of the cavity mode in its phase space and the state of the qubits, by the same angle $\phi$.
%Throughout this section, we set $\phi = 0$
%We consider an external driving 
Thus, we have $f(t) = \cos{(\omega t)} = (1/2)(e^{-i\omega t} + e^{i\omega t})$. Using this in Eq. (\ref{eq:He_rotframe}) and applying the RWA, we get
\begin{equation*}
	H_{e}^{\mathrm{RWA}} = \hbar\Omega\frac{1}{2}(a + a^{\dagger}).
\end{equation*}
Note that for the RWA to hold, the driving strength $\Omega$ should be small enough with respect to the driving frequency $\omega$.
The total Hamiltonian given in Eq. (\ref{eq:H}) becomes time-independent:
\begin{equation}\label{eq:H_RWA}
	H^{\mathrm{RWA}} = H_{g} + H_{e}^{\mathrm{RWA}}.
\end{equation}
The ground state of $H^{\mathrm{RWA}}$ is known for an arbitrary number of qubits \cite{Milburn}. It is normalizable when $\Omega < Ng$, where $N$ is the number of qubits. Let us denote the state vector corresponding to the ground state as $|E_{0};r\rangle$ with energy $E_{0}$. One can show that the ground state is given by a product state of the rotated qubits and a squeezed state of the cavity mode, with the ground-state energy $E_{0} = 0$. For our two-qubit case ($N=2$), it can be written as
\begin{equation*}
%	|E_{0}\rangle = R[\theta;\mathbf{e}_{y}] S(r) |00\rangle |0\rangle,
	|E_{0};r\rangle = |\theta_{r}\theta_{r}\rangle |r\rangle,
\end{equation*}
%The qubits are rotated from their ground state to become
where
\begin{equation}\label{eq:rotated_state_theta_r_theta_r}
	|\theta_{r}\theta_{r}\rangle \equiv R[\theta_{r};\mathbf{e}_{y}] |00\rangle
\end{equation}
and $R[\theta_{r};\mathbf{e}_{y}]$ rotates the qubits by an angle $\theta_{r}$ around the $y$-axis. The angle depends on the driving strength $\Omega$, as determined by the relation
\begin{equation*}
	\sin{\theta_{r}} = \frac{\Omega}{Ng} = \frac{\Omega}{2g},
\end{equation*}
where $0 \le \theta < \pi/2$ for $\Omega < Ng = 2g$.
The cavity mode is squeezed,
\begin{equation*}
	|r\rangle \equiv S(r)|0\rangle,
\end{equation*}
where $S(r) = \exp[(r/2)(a^{\dagger})^{2} - (r/2)a^{2})]$ is the squeezing operator with $r \ge 0$ being the squeezing parameter. The average photon number of $|r\rangle$ is
\begin{equation*}
	\bar{n}_{\gamma} \equiv \langle a^{\dagger}a \rangle = \sinh^{2}{r}.
\end{equation*}
The rotational angle is connected to the squeezing parameter by
\begin{equation}\label{eq:theta_r}
	\cos{\theta_{r}} = e^{-2r},
\end{equation}
where $0 \le r < \infty$. Figure \ref{fig:concurrence-along-squeezing}(a) illustrates the relation between $r$ and $\theta_{r}$. The average number of excited qubits is given as
\begin{equation*}
	\bar{n}_{q} \equiv \sum_{j=1}^{N} \langle \sigma_{j}^{+} \sigma_{j}^{-} \rangle = N \sin^{2}(\theta_{r}/2),
\end{equation*}
where $N=2$ is the number of qubits and the average is taken with respect to the rotated state, Eq. (\ref{eq:rotated_state_theta_r_theta_r}). The total average excitation number $\bar{n}$ is defined as the sum of the average number of photons and that of the excited qubits,
\begin{equation*}
	\bar{n} = \bar{n}_{\gamma} + \bar{n}_{q}.
\end{equation*}

\begin{figure}
	\centering
	\includegraphics[width=1\linewidth]{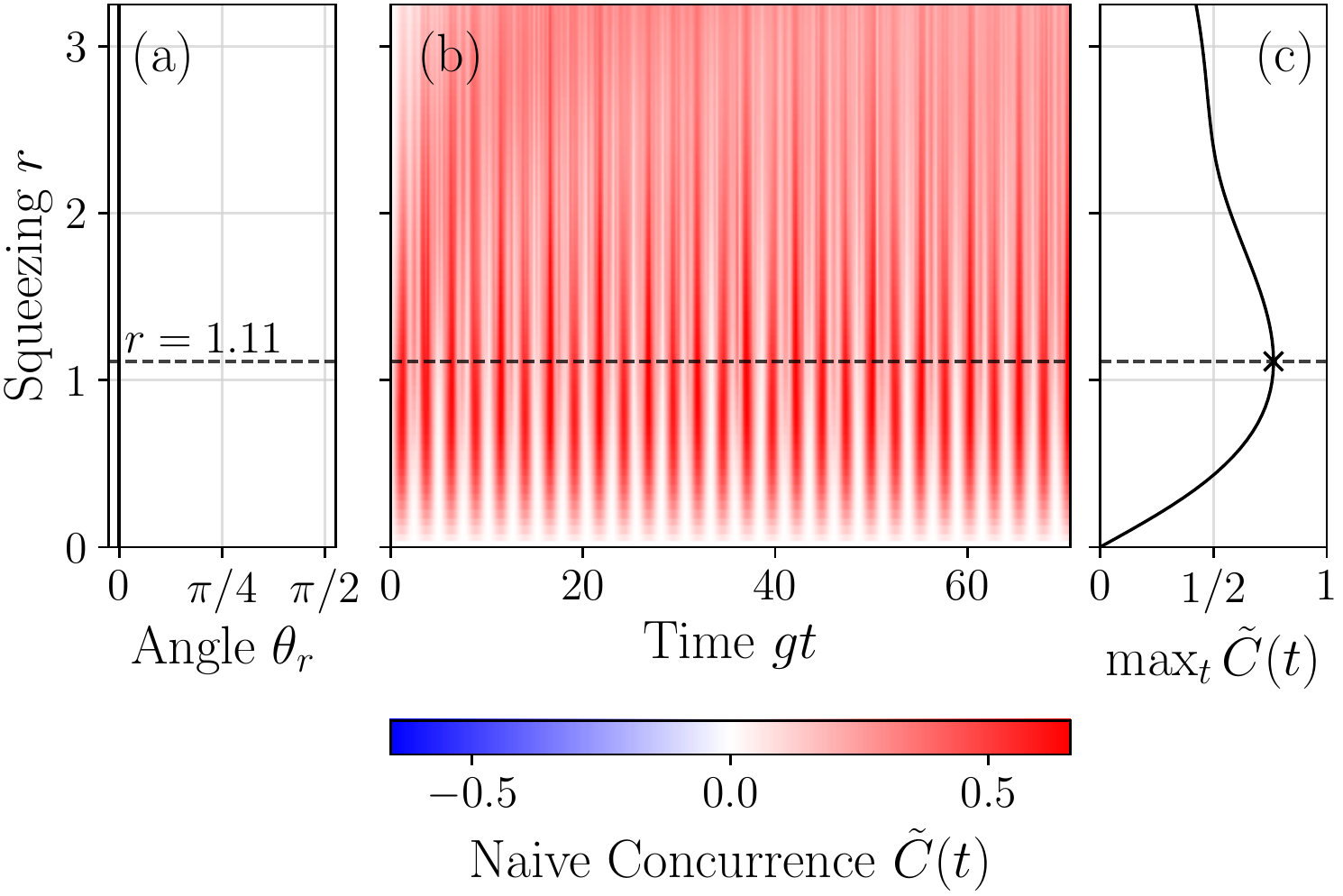}
	\caption{Same as Fig. \ref{fig:concurrence-along-squeezing}, except that the rotational angle is zero for all values of the squeezing parameter $r$.}
	\label{fig:concurrence-along-squeezing-with-no-rotation}
\end{figure}

By turning off the external driving at time $t = 0$, we mean $\Omega = 2g \sin{\theta_{r}}\rightarrow 0$ instantaneously. Then the system starts to evolve with the initial condition $|\psi(0)\rangle = |E_{0};r\rangle$ and with the Hamiltonian $H(t) = H_{g}$ since $\Omega = 0$.
%The concurrence depending on the squeezing parameter $r$ is shown 
The concurrence evolves accordingly for $t \ge 0$, which is shown in Fig. \ref{fig:concurrence-along-squeezing}(b).
In order to find the squeezing parameter with the maximal entanglement of formation, we evaluate the naive concurrence $\tilde{C}(t)$ maximized with respect to time, i.e. $\max_{t}{\tilde{C}(t)}$, which is shown in Fig. \ref{fig:concurrence-along-squeezing}(c). The maximization is done for the time range presented in Fig. \ref{fig:concurrence-along-squeezing}(b). The maximal concurrence occurs for $r = 0.899$ corresponding to the average photon number $\bar{n}_{\gamma} = 1.05$, average number of excited qubits $\bar{n}_{q} = 0.83$ and the average total excitation number $\bar{n} = 1.89$.
%To be accurate, the squeezing 
%which is set to be large enough to incorporate 
For larger squeezing, the maximal concurrence decreases. Note that the rotational angle converges to $\pi/2$ as $r \rightarrow \infty$.

To see the pure effect of the squeezed state on the generation of entanglement, one may rotate the qubits back to their ground states while keeping the squeezing of the cavity mode. For the rotations, one may address the qubits directly, by shining a laser along a direction perpendicular to the axis of the cavity. If it is not feasible to access the qubits directly, one can still rotate them by driving the cavity mode. In order to achieve this, one may drive the cavity mode with a specific pulse shape that satisfies $z(T) = 0$, in order to avoid displacement of the cavity mode but still to induce the required rotation of the qubits, facilitated by the second-order term discussed in Sec. \ref{sec:second-leading-order-effect-rotation}. An example of such a pulse shape can be found in Fig. \ref{fig:naive-concurr-rotation-dominant}(c).

In Fig. \ref{fig:concurrence-along-squeezing-with-no-rotation}(b), we show the naive concurrence induced purely by a squeezed state, without any rotation, i.e. $\theta_{r} = 0$, as shown in Fig. \ref{fig:concurrence-along-squeezing-with-no-rotation}(a). 
%Comparing the case where the rotation is involved which is shown in Fig. \ref{fig:concurrence-along-squeezing}(b), 
We notice that the naive concurrence can be negative when the rotation is involved as can be seen in Fig. \ref{fig:concurrence-along-squeezing}(b), whereas, when the rotation is subtracted out resulting in a pure squeezed state, the naive concurrence stays always nonnegative.
%We notice that the naive concurrence induced purely by a squeezed state is always nonnegative, whereas when the rotation remains

\begin{figure}
	\centering
	\includegraphics[width=1\linewidth]{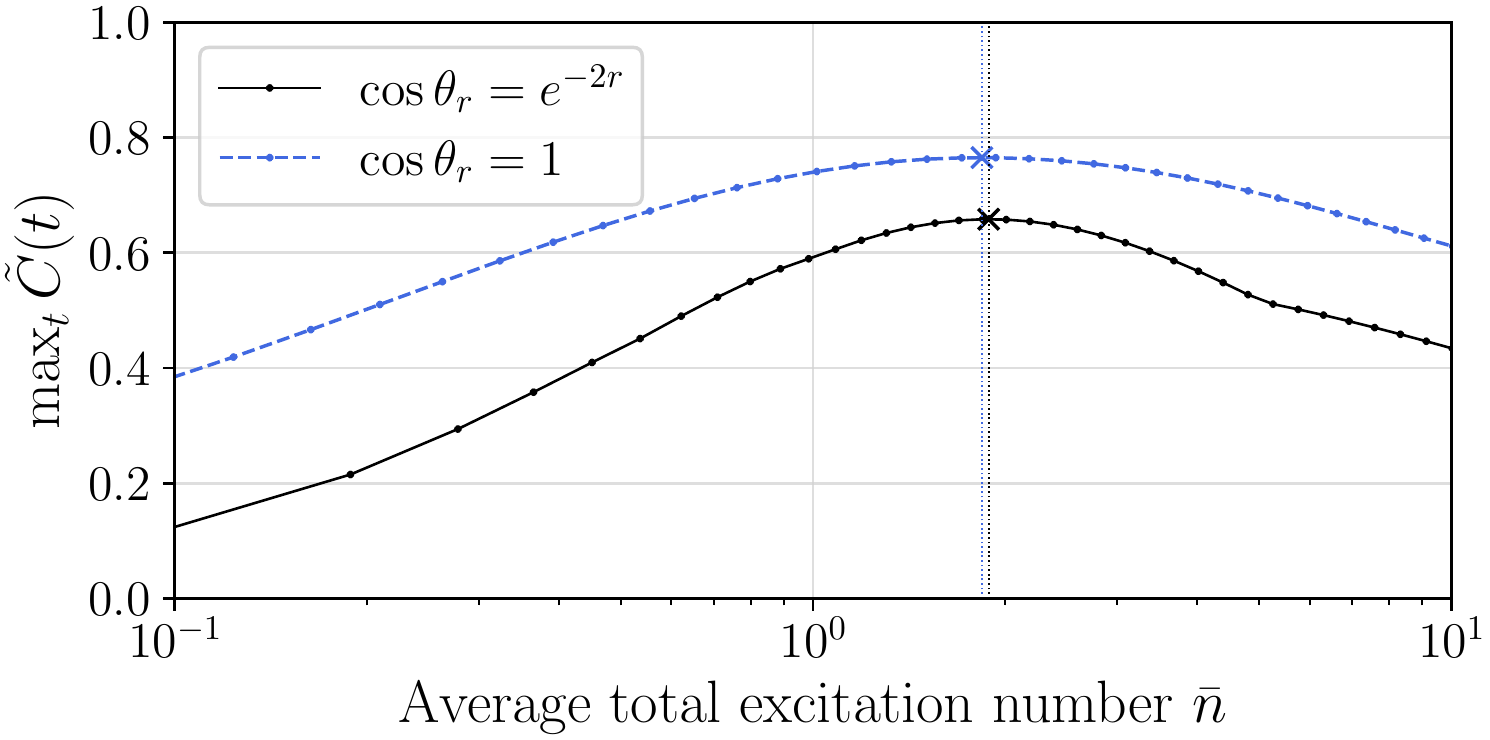}
	\caption{Naive concurrence maximized with respect to time versus the average total excitation number. The black solid line with circles denotes the case where the rotation is accompanied by the squeezing, whereas the blue dashed line represents the case where no rotation is involved.}
	\label{fig:naive-concurrence-max-versus-total-excitation-num}
\end{figure}

%What is also different between the two cases is that 
Another difference concerns the value of the squeezing parameter $r$ which maximizes the concurrence. As can be seen from Fig. \ref{fig:concurrence-along-squeezing-with-no-rotation}, the concurrence reaches its maximum at $r = 1.11$, which corresponds to the average photon number $\bar{n}_{\gamma} = 1.84$.
%The difference in the average photon number of the no-rotation case with that of rotated case is compensated by the rotation
%There is a difference between the average photon number of the 
The rotation shifts the optimal squeezing parameter $r$ and the corresponding average photon number. Without rotation, a higher average photon number is required to get the maximal concurrence. 
This difference in the average photon number is compensated by the pumping of quanta through the rotation, see Fig. \ref{fig:naive-concurrence-max-versus-total-excitation-num}.
%Since no rotation is involved, $\bar{n}_{q} = 0$. Thus, the total average excitation number is also given as $\bar{n} = \bar{n}_{\gamma} = 1.84$.
In order to generate a maximal concurrence, what matters most is the total number of excitations in the system, including both the photons and the excitations of the qubit system.
%The optimal total average excitation number

\begin{figure}
	\centering
	\includegraphics[width=1\linewidth]{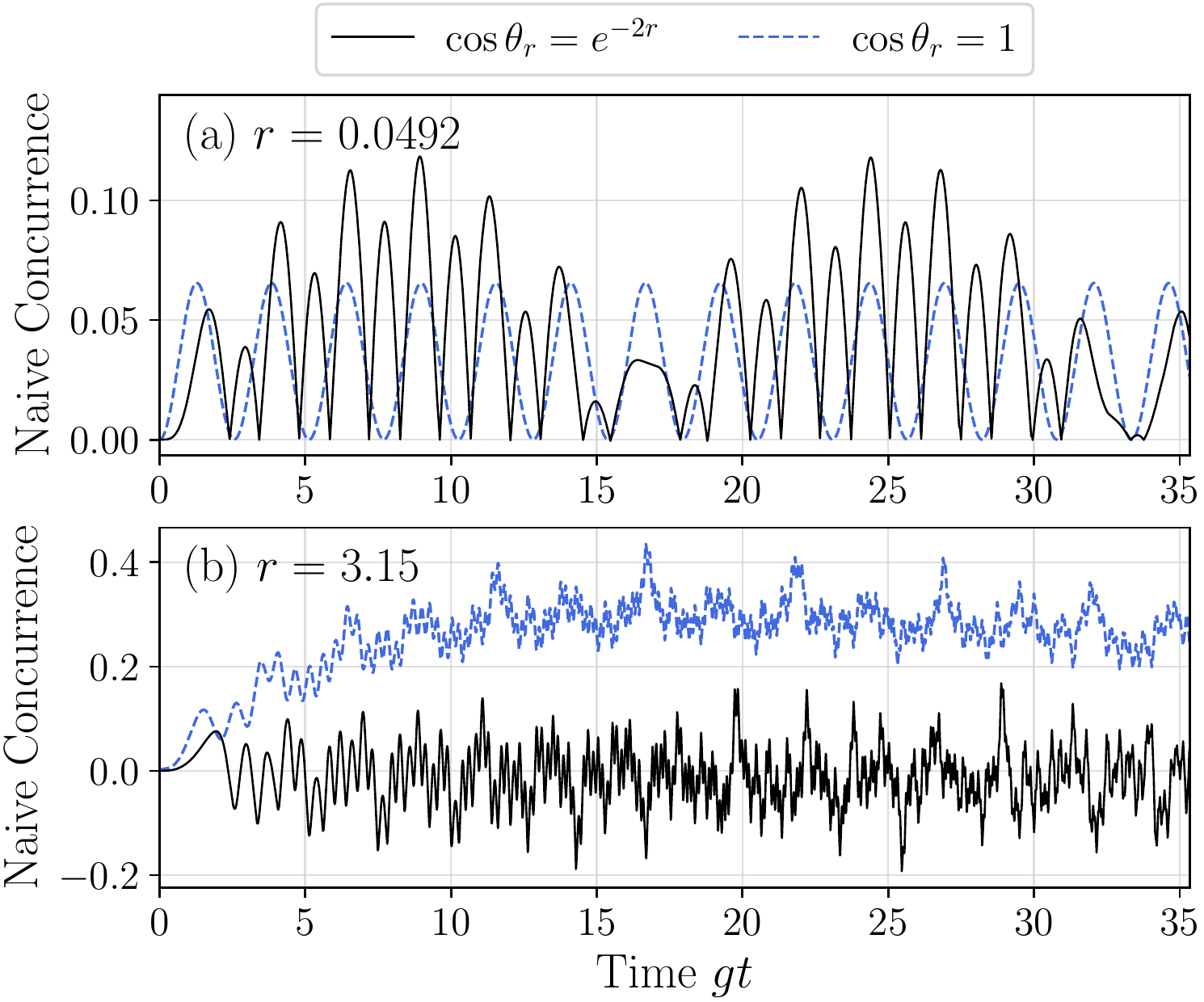}
	\caption{
		Comparison of the naive concurrence with and without the rotation. The squeezing parameters are shown in each panel. The black solid lines represent the inducing of concurrence by a squeezed state with no rotation.
		%		The black solid lines represent the time-dependent naive concurrence with an initial state which is squeezed and has zero rotation
		%		with a squeezed state $|r\rangle$ as the initial state and with zero rotational angle $\theta_{r}=0$. 
		The blue dashed lines represent the naive concurrence from a squeezed and rotated state with the rotational angle $\theta_{r}$ given as Eq. (\ref{eq:theta_r}).
	}
	\label{fig:squeezed-state-naive-concurr-comparision}
\end{figure}

We finish this section by considering two limits. The first is the low-excitation limit, $\bar{n}\ll 1$. In Fig. \ref{fig:squeezed-state-naive-concurr-comparision}(a), we show the time-dependent naive concurrence in the low-squeezing regime, where $r = 0.0492$ corresponding to $\bar{n} = 0.0962$ with rotation and $\bar{n} = 0.00243$ without rotation. We see that when there is no rotation, an oscillation with a well-defined period is present. The reason is that in this limit the initial state consists of $|00;0\rangle$ and a small amount of $|00;2\rangle$. The former has no time dependence. The latter belongs to the two-quanta subspace and is essentially the only contribution to the time dependence in this regime. When the rotation enters, however, the oscillation of the concurrence has multiple frequencies, as can be seen in Fig. \ref{fig:squeezed-state-naive-concurr-comparision}(a), indicated by the black solid line. This can be understood as a consequence of an additional interference with another state with a single quantum, namely $|\Psi^{+};0\rangle$, introduced by the rotation of $|00;0\rangle$.
%Since $|\Psi^{+};0\rangle$ belongs to another subspace, the one with a single quantum, the interferences among the zero-, single- and the two-quanta state exist in addition to that between the zero- and two-quanta state.
%the dynamics in essentially determined by the interference between the 
%When the rotation enters represented by the black solid line in Fig. \ref{fig:squeezed-state-naive-concurr-comparision}(a),
%the vacuum state and a small amount of two-photon state. The vacuum state belongs

We next turn to the opposite regime, where the excitation number is much higher than $1$. One example is shown in Fig. \ref{fig:squeezed-state-naive-concurr-comparision}(b), where the squeezing parameter is $r = 3.15$. The corresponding average total excitation number with rotation is $\bar{n} = 137$ and that without rotation is $\bar{n} = 136$. After some time passes, the naive concurrence in both cases shows complicated fluctuations. When the rotation is not subtracted, the naive concurrence fluctuates around an average value close to zero, whereas when the initial state is purely a squeezed state with the rotation removed, the system can maintain positive entanglement of formation for longer duration.

\section{\label{sec:discussion}Discussion}

Let us discuss the relevant parameters for experimental realization of the presented results. Firstly, since the theory employs the RWA for the coupling between each qubit and the cavity mode, the coupling strength should be small compared to the frequency of the cavity mode and the qubits, i.e. $g \ll \omega$. Secondly, for the subcycle, or sub-Rabi, driving, we require $g\tau_{d} \ll 1$, which follows from $\tau_{d} \ll T_{g} = (\sqrt{n}g)^{-1}$. For the efficient coupling of the external field to the cavity mode, without affecting much the other modes, the pulse needs to have a well-defined carrier frequency which is resonant with the frequency of the cavity mode. From this condition, we require $\omega\tau_{d} \gg 1$, which follows from $T_{\omega} \equiv 2\pi / \omega \ll \tau_{d}$. All the conditions can be summarized by Eq. (\ref{eq:pulse-duration-condition}). As long as the cavity-qubit coupling $g$, the mode frequency $\omega$ and the pulse duration $\tau_{d}$ satisfies this condition, one can test the demonstrated results. To have a concrete example, we consider a quantum dot in a photonic crystal \cite{Faraon2008}, where the resonant frequency corresponds to a wavelength $\lambda = 928\,\mathrm{nm}$ and the cavity-qubit coupling $g/2\pi = 16\,\mathrm{GHz}$. This imposes a condition on the pulse duration as 
\begin{equation*}
	T_{\omega} \equiv 2\pi/\omega \sim 3\,\mathrm{fs} \ll \tau_{d} \ll 55\,\mathrm{ps} / \sqrt{n} \sim 2\pi/(\sqrt{n}g).
\end{equation*}
%$T_{\omega} = 2\pi/\omega \sim 3 \mathrm{fs} \ll \tau_{d} \ll 55 \mathrm{ps} / \sqrt{n} \sim 2\pi/(g\sqrt{n})$.
%
For the low-excitation limit, where $n \sim 1$, we get $\tau_{d} \ll 55\,\mathrm{ps}$. If one selects $\tau_{d} = 5.5\,\mathrm{ps}$, it corresponds to $g\tau_{d} \sim 0.6$, which is used in our calculations, e.g. in Fig. \ref{fig:fidel-comparison-for-different-pulse-shapes}.

%\begin{itemize}
%	\item The nature of the coupling - is it like a beam splitter? Basically, where does the interaction Hamiltonian of the form $\hbar \Omega_{e} f(t) x$ come from?
%	\item Validity of the model for the impulsive pulse - issues on exciting other cavity modes - is it possible to justify it?
%	\begin{itemize}
%		\item Look into each cavity system, especially have a look at the photonic crystal and check if other modes within the bandwidth of the driving pulse are forbidden to be excited.
%		\item Try to find a specific value of the cavity frequency $\omega$ and the applicable $\Omega_{e}$ and $\tau_{d}$ as well as the coupling $g$.
%	\end{itemize}
%	\item The coupling strength 
%	\item Propose a set of parameters for a possible experimental realization.
%	\item (TODO) Cite the paper which Shaul recommended.
%\end{itemize}

\section{\label{sec:conclusion}Conclusions}

We have considered the generation of entanglement between two qubits by using a classical light source and a quantized cavity mode. We have shown how two qubits can be entangled by exchanging quanta with a third party which in our case is the cavity mode. Quanta can be pumped into the system through an external driving by a classical light source coupled to the cavity mode, with no direct driving of the qubits.
The quanta exchange timescale $T_{g} = (\sqrt{n}g)^{-1}$ is identified. With respect to this characteristic timescale of the cavity-qubits system, we considered two regimes of the external driving. 

We first discussed the subcycle driving, where it is performed by a pulse with duration shorter than the characteristic timescale of the system, $T_{g}$. We showed that the leading-order effect of a pulsed driving is a displacement of the cavity mode, which can be expected since the cavity mode is directly coupled to the pulse. We further showed that by shaping the pulse, one can also rotate the qubits, and if desired, one can let the cavity remain intact after the passage of the pulse. The entanglement generation for each type of the pulse shape was demonstrated, showing good agreement with exact results. We showed that the error for the displacement operation can be set arbitrarily small by choosing a sufficiently small $g\tau_{d}$, which represents the shortness of the pulse duration with respect to $T_{g}$.
The error was estimated by identifying the convergence rate. Furthermore, enhancing the convergence rate by shaping the pulse was demonstrated, indicating how to perform a desired operation with a given fidelity. Higher-order effects including the phase shift of the qubits and the displacement of the cavity mode conditional to the qubits state are found.

As the opposite regime of the driving, we discussed a quasistatic driving where its duration is much longer than $T_{g}$. We considered a continuous-wave driving with the driving amplitude such that there exists a normalizable ground state in the rotating frame. In this regime, the ground state is a squeezed state with rotated qubits. Assuming adiabatic driving to prepare the ground state with nonzero squeezing, we studied the entanglement induced by the squeezed and rotated state. We observed a maximal entanglement of formation when the total number of excitations, which is a sum of the average photon number and the average number of excited qubits, is on the order of 1. We compared the result with the case of pure squeezing where there is no rotation of the qubits and found that the optimal value of the squeezing parameter slightly changes. However, the average total number of excitations which generates the maximal entanglement was found to remain essentially the same.
%We also find that almost same number of total excitations were needed to obtain the maximal entanglement of formation when there is only a squeezed state without rotation.
%We showed this is also true when the rotation is removed.

The studied cavity-qubits system is a useful testbed for fundamental quantum properties of light-matter interaction and entanglement. 
%Being able to control the state of the system is essential for an  experimental verification of any prediction of the quantum theory. 
The presented framework enables selecting specific operations on the joint cavity-qubits state by an appropriate pulse shaping of an external classical light. The set of all possible operations accessible by the subcycle or the quasistatic driving and how each operation can be activated or suppressed with prescription for a high fidelity 
%would help the experimental design.
%based on a coherent light and a cavity. 
%This 
can be used for a laser-based experimental generation and control of the entanglement between non-interacting systems. 
\appendix

%\section{Evaluation of $U_{g}(t)$}\label{sec:Ug-and-fock-state-evolution}
%(expression of $U_{g}(t)$)
%(Derivation of Fock states)
%(Evaluation of the reduced density matrix driven by an subcycle pulse)

%\section{(derivation of time-evolution operator in the expansion in terms of $\tilde{\tau}_{d}$)}\label{sec:Un-deriv}
%(derivation goes here)

%\section{}
%
%\begin{equation}
%\begin{split}
%	\tilde{H}^{\prime}_{II}(u) 
%	& \equiv \frac{\tau_{d}}{\hbar}H^{\prime}_{II}(u)\\
%	& \equiv \sum_{\mu=1}^{\infty}\sum_{\nu=0}^{\infty}{
%		(\Omega\tau_{d})^{\mu} (g\tau_{d})^{\mu+\nu} \tilde{H}_{II}^{\prime(\mu,\mu+\nu)}(u)
%	}
%\end{split}
%\end{equation}
%
%\begin{equation*}
%	\tilde{H}_{II}^{\prime(j,k)}(u) 
%	\equiv \frac{u^{k}}{k!} f(u) x_{\omega}^{(j-1,k)}(u)
%\end{equation*}
%
%\begin{equation*}
%	x_{\omega}^{(j-1,k)}(u)
%	\equiv \sum_{\substack{j_{1}+\cdots+j_{k-1} = j-1 \\ j_{1},\cdots,j_{k-1} \in \{0,1\}}}
%	{
%		\mathrm{ad}_{i\tilde{H}_{g,j_{1}}}\cdots \mathrm{ad}_{i\tilde{H}_{g,j_{k-1}}} \mathrm{ad}_{i\tilde{H}_{g,0}} x_{\omega}(u)
%	}
%\end{equation*}

\section{\label{sec:validity-of-cavity-qubit-rwa}Validity of the RWA for a short pulse}
\begin{figure}
	\centering
	\includegraphics[width=\linewidth]{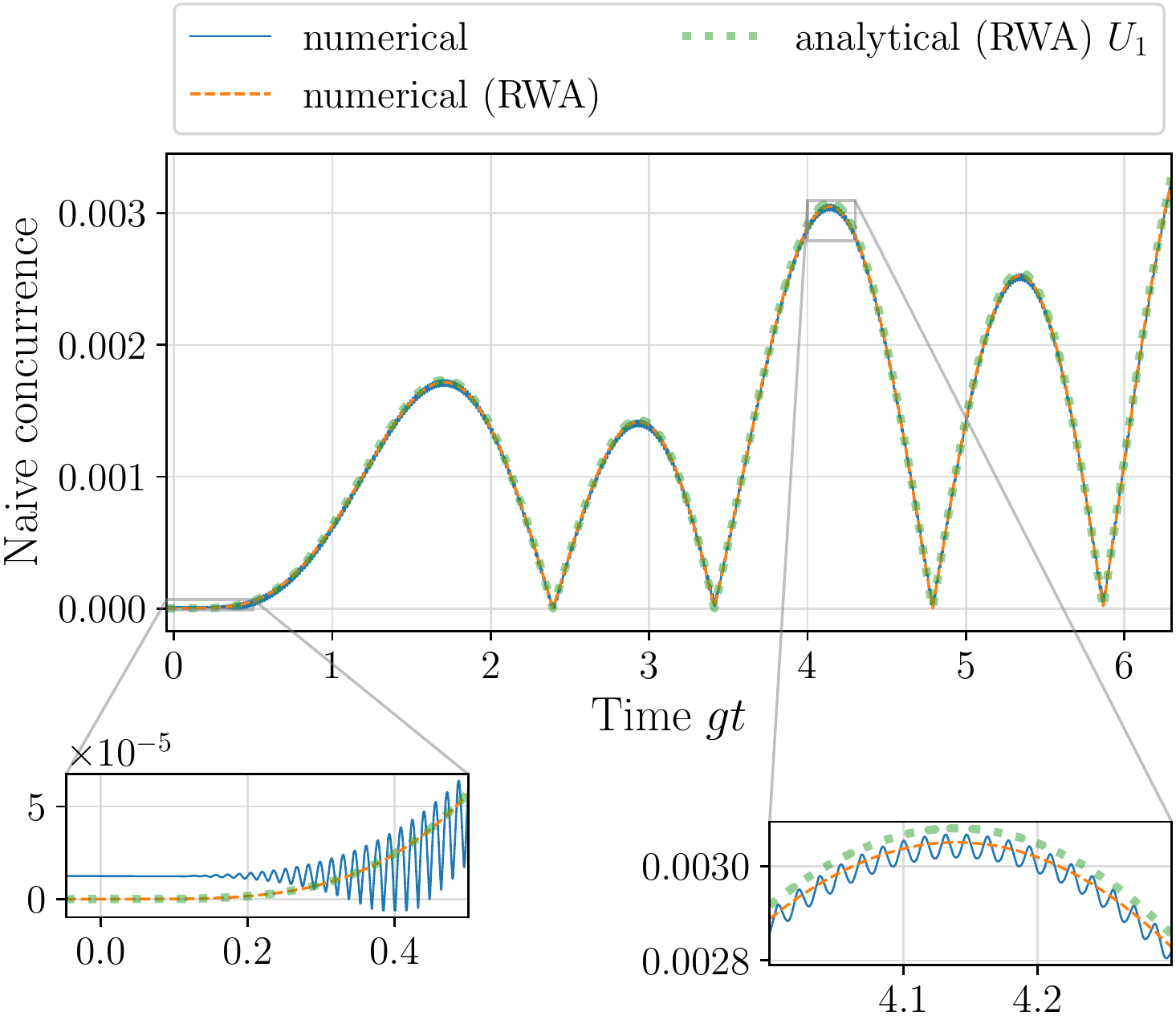}
	\caption{The naive concurrence of the two qubits calculated with different methods. The dashed and solid lines are numerical solutions, respectively with and without the rotating wave approximation (RWA) between the cavity and each qubit. The dotted line is obtained by the analytical expression based on $\mathcal{U} \simeq U_{1}$.
%		The dashed line is also evaluated numerically 
	The cavity mode is driven by a pulse with duration $\tau_{d} = T_{\omega} / 45$, strength $\Omega = 0.225\omega$ and  shape $f(t) = \exp[-(t/\tau_{d})^{2}]$. The coupling strength between the cavity mode and each qubit is $g = 0.005\omega$.}
	\label{fig:naive-concurrence-and-lambda1-analytical-test-049}
\end{figure}
% e^{-t^{2}/\tau_{d}^{2}

We check the validity of the usage of the RWA in Eq. (\ref{eq:Hg}) on the coupling between the cavity mode and each qubit. In Fig. \ref{fig:naive-concurrence-and-lambda1-analytical-test-049}, we compare the naive concurrence obtained with and without the RWA. 
In each calculation, their respective ground states are used. Note that the ground state under the RWA is $|00;0\rangle$ which is separable, whereas the ground state without the RWA is entangled with concurrence on the order of $10^{-5}$.
In the full (RWA-free) result, there is a relatively rapid oscillation on top of the longer-scale evolution, coming from the counter-rotating terms in the Hamiltonian, which are neglected under the RWA. If the amplitude of the rapid oscillation becomes comparable to the magnitude of the naive concurrence, one may not completely ignore the counter-rotating terms. In the studied cases the value of the naive concurrence is large enough with respect to the rapid oscillation, justifying the RWA.
%When the naive concurrence is large enough than the rapid oscillation, one can see that the RWA is a good approximation. 
We note that all data presented in this paper do not rely on the RWA between the cavity mode and the external field, being consistent with Eq. (\ref{eq:He_rotframe}).
%When the naive concurrence is of the same order as the amplitude of the oscillation 

\section{\label{sec:magnus-expansion-order-of-magnitudes}Orders of magnitudes of Magnus terms}
%Convergence of Magnus expansions

In this section, we derive Eqs. (\ref{eq:A_I_m_t}) and (\ref{eq:A_II_prime_m_t_in_tilde_Omega_n_tilde_g_n}) to obtain sufficient conditions for the convergence of the Magnus expansions in Eq. (\ref{eq:magnus-exponent}) and (\ref{eq:A_II_t_magnus}), respectively. For general discussions on the Magnus expansion, see Ref. \cite{Magnus}.

Let us start from the first Magnus expansion, Eq. (\ref{eq:A_I_m_t}). 
%where each term can be written as
%\begin{equation*}
%\end{equation*}
The first term of the expansion can be written as 
\begin{equation}\label{eq:A_I_1_t_and_tilde_A_I_1_t}
	A_{I}^{(1)}(t) \equiv (\Omega\tau_{d}) \tilde{A}_{I}^{(1)}(t),
\end{equation}
%$A_{I}^{(1)}(t) \equiv (\Omega\tau_{d}) \tilde{A}_{I}^{(1)}(t)$, 
where
\begin{equation}\label{eq:tilde_A_I_1_t}
\begin{split}
	\tilde{A}_{I}^{(1)}(t) 
	& = \int_{-T/\tau_{d}}^{t/\tau_{d}}{du\,}{ \tilde{H}_{I}(u) }\\
	& = \mathcal{O}[(\Omega\tau_{d})^{0}]
\end{split}
\end{equation}
and $\tilde{H}_{I}(u) \equiv {H}_{I}(u) / \hbar \Omega$ is defined in Eq. (\ref{eq:H_I_t}). Any following term, i.e. $A_{I}^{(m)}(t)$ for $m > 1$, can be written in terms of its preceding terms, i.e. $A_{I}^{(k)}(t)$ for $1 \le k < m$, as \cite{BLANES2009151}
\begin{widetext}
\begin{equation}\label{eq:A_I_m_t_magnus_explicit}
	A_{I}^{(m)}(t) 
	= \Omega\tau_{d} \sum_{j=1}^{m-1} \frac{B_{j}}{j!}
	\sum_{
		\substack{
			k_{1}+\cdots+k_{j} = m-1 \\ 
			k_{1} \ge 1,\cdots,k_{j} \ge 1
		}
	}
	\int_{-T/\tau_{d}}^{t/\tau_{d}}{
		du\,
		\mathrm{ad}_{ -i A_{I}^{(k_{1})}(u) }
		\cdots
		\mathrm{ad}_{ -i A_{I}^{(k_{j})}(u) }
		\tilde{H}_{I}(u)
	}.
\end{equation}
\end{widetext}
$B_{j}$ for a nonnegative integer $j$ is the Bernoulli number \cite{bernoulli1713ars,arfken} and $\mathrm{ad}_{X}{Y} \equiv [X,Y]$ for given operators $X$ and $Y$. For example, the second Magnus term is given by
\begin{equation*}
	A_{I}^{(2)}(t) = \Omega\tau_{d} \left(-\frac{1}{2}\right) \int_{-T/\tau_{d}}^{t/\tau_{d}} 
	du\,[-iA_{I}^{(1)}(u), \tilde{H}_{I}(u)],
\end{equation*}
with $B_{1} = -1/2$.
%The first two terms read
%\begin{equation*}
%\begin{split}
%	A_{I}
%\end{split}
%\end{equation*}

From Eq. (\ref{eq:A_I_m_t_magnus_explicit}), let us show 
\begin{equation}\label{eq:A_I_m_t_order_of_mag_in_Omega_tau_d}
\begin{split}
	{A}_{I}^{(m)}(t) 
	& \equiv (\Omega\tau_{d})^{m} \tilde{A}_{I}^{(m)}(t)\\
	& = \mathcal{O}[(\Omega\tau_{d})^{m}]
\end{split}
\end{equation}
%$\tilde{A}_{I}^{(m)}(t) = \mathcal{O}[(\Omega\tau_{d})^{m}]$ 
for all $m \ge 1$, by an induction. 
This implies $\tilde{A}_{I}^{(m)}(t) = \mathcal{O}[(\Omega\tau_{d})^{0}]$ for all $m$.
Equation (\ref{eq:A_I_m_t_order_of_mag_in_Omega_tau_d}) holds for $m=1$, which follows from Eqs. (\ref{eq:A_I_1_t_and_tilde_A_I_1_t}) and (\ref{eq:tilde_A_I_1_t}). For any $m > 1$, if Eq. (\ref{eq:A_I_m_t_order_of_mag_in_Omega_tau_d}) holds for all $k$ such that $1 \le k < m$, we set ${A}_{I}^{(k)}(t) \equiv (\Omega\tau_{d})^{k} \tilde{A}_{I}^{(k)}(t)$ and use $\mathrm{ad}_{cX}{Y} = c\,\mathrm{ad}_{X}{Y}$ for $c \in \mathbb{C}$ to show that Eq. (\ref{eq:A_I_m_t_magnus_explicit}) is proportional to $(\Omega\tau_{d})^{m}$. This concludes the induction for Eq. (\ref{eq:A_I_m_t_order_of_mag_in_Omega_tau_d}) to hold for all $m \ge 1$.

We then proceed to show
\begin{equation}\label{eq:tilde_A_I_m_t_sqrt_n_to_m}
	\tilde{A}_{I}^{(m)}(t) = \mathcal{O}[(\sqrt{n})^{m}],
\end{equation}
for all $m \ge 1$, where $n$ is the number of excitations in the state of the system. 
%on which the time-evolution operator $\mathcal{U}$ acts
When the state is in a superposition of states with different numbers of excitations, $n$ may be set to the average number of excitations. Each $\sqrt{n}$ comes from $a$ or $a^{\dagger}$. Equation (\ref{eq:tilde_A_I_m_t_sqrt_n_to_m}) holds for $m=1$, which follows from Eq. (\ref{eq:tilde_A_I_1_t}) and the fact that $\tilde{H}_{I}(u)$ is linear in $a$ and $a^{\dagger}$ in its leading order, as can be seen from Eqs. (\ref{eq:tilde_H_I_g-tau_d-expansion}) and (\ref{eq:x_omega_t}). Showing Eq. (\ref{eq:tilde_A_I_m_t_sqrt_n_to_m}) for any $m > 1$ can be done by another induction in the same manner as we did for deriving Eq. (\ref{eq:A_I_m_t_order_of_mag_in_Omega_tau_d}).

Combining Eq. (\ref{eq:A_I_m_t_order_of_mag_in_Omega_tau_d}) with (\ref{eq:tilde_A_I_m_t_sqrt_n_to_m}), we obtain Eq. (\ref{eq:A_I_m_t}). Similarly, Eq. (\ref{eq:A_II_prime_m_t_in_tilde_Omega_n_tilde_g_n}) can be derived by substituting $A_{I}^{(k)}(t)$ and $\tilde{H}_{I}(u)$ in Eq. (\ref{eq:A_I_m_t_magnus_explicit}) for $A_{II}^{\prime(k)}(t)$ and $\tilde{H}_{II}^{\prime}(t)$, respectively, for all $k$ such that $1 \le k \le m$.

%(compare the two Magnus expansions - for $H_{I}(t)$ and $H_{II}^{\prime}(t)$)

%(Show that $A_{I}^{(m)}(t) = \mathcal{O}[(\sqrt{n}\Omega\tau_{d})^{m}]$ and $A_{II}^{\prime(m)}(t) = \mathcal{O}[(\sqrt{n}\Omega\tau_{d})^{m}(\sqrt{n}g\tau_{d})^{m}]$)

%
%\section{Expression of concurrence from a coherent state}\label{sec:appendix-concurr-expr-from-coh-state}
%(present the expression of the concurrence with some brief derivation.)

\begin{acknowledgments}

S.A. was supported by the education and training program of the Quantum Information Research Support Center, funded through the National research foundation of Korea (NRF) by the Ministry of science and ICT (MSIT) of the Korean government under number 2021M3H3A103657313. 
S.A. and A.S.M. were supported by the National Research Foundation of Korea (NRF) grant funded by the Korean government (MSIT) under number 2020R1A2C1008500.
V.Y.C. and S.M. were supported by the U.S. Department of Energy (DOE), Office of Science, Basic Energy Sciences, under Award Number DE-SC0022134. S.M. was also supported by the National Science Foundation (NSF).

S.A. is grateful to God, who created the Heaven and the Earth, to be faithful in helping him whenever asked for wisdom and the way to go for this research.

%(ACKNOWLEDGMENTS goes here.)
%\begin{itemize}
%	\item Discussion with Dr. Bing Gu on the validity of this model as well as general discussions regarding the project. Pointing out the entanglement formed by the counter-rotating terms and the necessity of checking the validity of the RWA.
%	\item (Discussion with Hari on the collapse and revival)
%	\item (Discussion with Dr. Sandeep Salma on the collapse and revival, along with the Schr\"odinger's cat state.)
%	\item relevant grants which include
%	\begin{itemize}
%		\item "This research was supported by the education and training program of the	Quantum Information Research Support Center, funded through the National research foundation of Korea(NRF) by the Ministry of science and ICT(MSIT) of the Korean government(No.2021M3H3A103657313)."
%		\item Other relevant grants from the authors. (See the abstract of OSK conference)
%	\end{itemize}
%	\item S.A. is grateful to God, the Creator of the Heaven and the Earth, for leading his way on this project and giving wisdom to him in proceeding this work.
%\end{itemize}
\end{acknowledgments}

%\bibliography{manuscript}



\end{document}